\documentclass[letterpaper,aps,prd,twocolumn,tightenlines,preprintnumbers,nofootinbib,showkeys,superscriptaddress,longbibliography]{revtex4-1}

\usepackage{lipsum} 

\usepackage{XCharter}
\usepackage[T1]{fontenc}
\usepackage{bm}
\usepackage{bbold}
\usepackage{pifont}
\usepackage{rotating}
\usepackage{fullpage}
\usepackage{amsfonts,shorthand}
\usepackage{amsmath}
\usepackage{slashed}
\usepackage{siunitx}
\usepackage{amssymb}
\usepackage{graphicx}
\usepackage{epic}
\usepackage{eepic}
\usepackage{epsfig}
\usepackage{latexsym}
\usepackage[dvipsnames]{xcolor}
\usepackage[export]{adjustbox}
\usepackage{float}
\usepackage{multirow, makecell}
\usepackage{hyperref}
\usepackage{enumitem}
\hypersetup{colorlinks=true,citecolor=red,linkcolor=NavyBlue,urlcolor=NavyBlue}
\usepackage[caption=false]{subfig}
 
\usepackage{natbib}
\usepackage{relsize}
\usepackage{wrapfig}
\usepackage[left=1.65cm,right=1.65cm,top=1.83cm,bottom=1.84cm]{geometry}
\usepackage{slashed}

\begin{document}
\relscale{1.05}
\title{Tagging fully hadronic exotic decays of the vectorlike $\mathbf{B}$ quark using a graph neural network}
\author{Jai Bardhan}
\email{jai.bardhan@research.iiit.ac.in} 
\affiliation{Center for Computational Natural Sciences and Bioinformatics, International Institute of Information Technology, Hyderabad 500 032, India}

\author{Tanumoy Mandal}
\email{tanumoy@iisertvm.ac.in}
\affiliation{Indian Institute of Science Education and Research Thiruvananthapuram, Vithura, Kerala, 695 551, India}

\author{Subhadip Mitra}
\email{subhadip.mitra@iiit.ac.in}
\affiliation{Center for Computational Natural Sciences and Bioinformatics, International Institute of Information Technology, Hyderabad 500 032, India}
\affiliation{Center for Quantum Science and Technology, International Institute of Information Technology, Hyderabad 500 032, India}

\author{Cyrin Neeraj}
\email{cyrin.neeraj@research.iiit.ac.in}
\affiliation{Center for Computational Natural Sciences and Bioinformatics, International Institute of Information Technology, Hyderabad 500 032, India}

\author{Mihir Rawat}
\email{mihir.r@research.iiit.ac.in}
\affiliation{Center for Computational Natural Sciences and Bioinformatics, International Institute of Information Technology, Hyderabad 500 032, India}
\date{\today}
\begin{abstract}
\noindent 
Following up on our earlier study in [J. Bardhan \emph{et al.}, Machine learning-enhanced search for a vectorlike singlet B quark decaying to a singlet scalar or pseudoscalar, \href{https://doi.org/10.1103/PhysRevD.107.115001}{{\it Phys. Rev.} D {\bf 107} (2023) 115001}; \href{http://arxiv.org/abs/2212.02442}{arXiv:2212.02442 [hep-ph]}], we investigate the LHC prospects of pair-produced vectorlike $B$ quarks decaying exotically to a new gauge-singlet (pseudo)scalar field $\Phi$ and a $b$ quark. After the electroweak symmetry breaking, the $\Phi$ decays predominantly to $gg/bb$ final states, leading to a fully hadronic $2b+4j$ or $6b$ signature. Because of the large Standard Model background and the lack of leptonic handles, it is a difficult channel to probe. To overcome the challenge, we employ a hybrid deep learning model containing a graph neural network followed by a deep neural network. We estimate that such a state-of-the-art deep learning analysis pipeline can lead to a performance comparable to that in the semi-leptonic mode, taking the discovery (exclusion) reach up to about $M_B=1.8\:(2.4)$~TeV at HL-LHC when $B$ decays fully exotically, i.e., BR$(B \to b\Phi) = 100\%$.

\end{abstract}
\maketitle

\section{Introduction}
\noindent 
New fermions are usually considered vectorlike in scenarios beyond the Standard Model (SM) to avoid introducing gauge anomalies~\cite{Gopalakrishna:2011ef,Gopalakrishna:2013hua,Alves:2023ufm}. The LHC has been searching for vectorlike quarks (VLQs) that mix with the third-generation ones at the ATLAS and CMS detectors (see, e.g., Ref.~\cite{CMS:2024bni}). 
So far, none of these detectors has observed any VLQ signatures, pushing the mass exclusion limits on them to about $2$ TeV. Recently, there has been considerable interest in non-standard decays of VLQs ~\cite{Serra:2015xfa,Arhrib:2016rlj,Han:2018hcu,Bizot:2018tds,Benbrik:2019zdp,Cacciapaglia:2019bqz,Aguilar-Saavedra:2019ghg,Cacciapaglia:2019zmj,Wang:2020ips,Banerjee:2022xmu,Banerjee:2022izw,Elander:2023aow,Franceschini:2023nlp,Banerjee:2023upj,Bennett:2023rsl,Banerjee:2024zvg,Qureshi:2024naw}. While there could be various motivations to add extra particles to the new particle spectrum from a model-building perspective (consider, e.g., the Maverick Top-partner model~\cite{Kim:2019oyh}, where the VLQ can decay to a dark photon~\cite{Verma:2022nyd,Verma:2024kdx}), any extra decay mode would reduce the branching ratios (BRs) in the standard modes. Since the experiments considered only the standard modes, the limits on VLQs could be actually weaker.

This paper is part of a series (see Refs.~\cite{Bhardwaj:2022nko,Bhardwaj:2022wfz,Bardhan:2022sif}) where we study the prospects in looking for VLQs decaying through an exotic decay mode, $q\Phi$ (where $q$ is a third-generation quark and $\Phi$ is a new spinless boson singlet under the SM gauge group), at the LHC. In Ref.~\cite{Bhardwaj:2022nko}, we charted out the possibilities at the LHC. Since a spinless gauge-singlet particle does not couple to SM particles directly (except, possibly, the Higgs), it is difficult to probe experimentally. Hence, a priori, it might seem that a VLQ dominantly decaying to $\Phi$ would have poor prospects at the LHC; however, in Refs. \cite{Bhardwaj:2022wfz,Bardhan:2022sif}, we showed that this is not the case. Even if, for a conservative estimation, we assume $\Phi$ has no direct coupling to particles other than a VLQ, it gains such couplings once the electroweak symmetry (EWS) breaks and the VLQ mixes with SM quarks. Depending on the VLQ(s), a $\Phi$ could then decay to a pair of $b$ or $t$ quarks at the tree level and to a pair of SM bosons (like $gg$, $\gamma\gamma$, $Z\gamma$ or a pair of heavy SM bosons if kinematically allowed) at the loop level. Interestingly, the loop-mediated decay of the $\Phi$ to a pair of gluons can dominate over all other decay modes, including the tree-level ones, in a large part of the parameter space. We analysed the $pp \to TT \to \left(t \Phi\right)\,\left(t\Phi\right)\to (tgg)(tgg) $ process to estimate the discover/exclusion prospects of a singlet $T$ quark at the LHC in Ref.~\cite{Bhardwaj:2022wfz}. We studied the $B$ quark prospects in Ref.~\cite{Bardhan:2022sif} through a mixed-decay channel: $pp \to B B \to \left(b \Phi\right)\,\left(tW\right)\to(bgg/bbb)(tW)$. Our analyses showed good prospects for both cases, especially with machine learning techniques.

\begin{figure}[!t]
    \centering
    \includegraphics[width=0.375\textwidth]{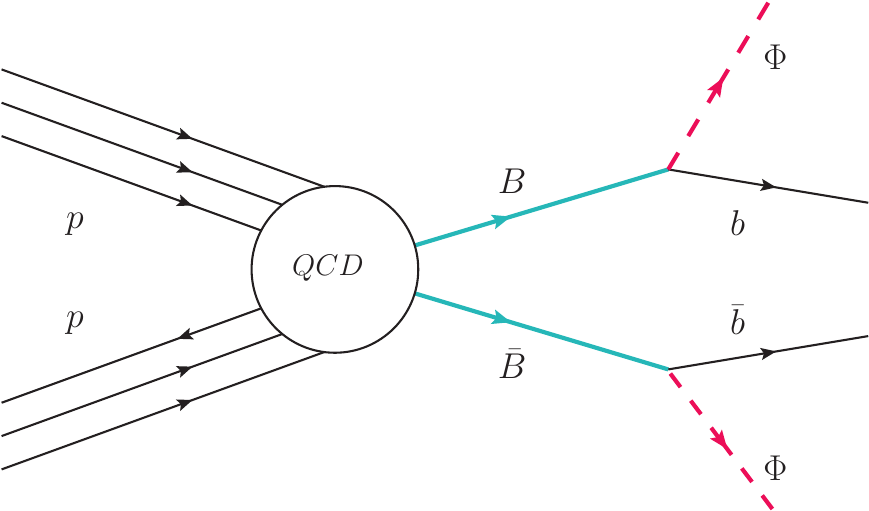}
    \caption{Signal topology}
    \label{fig:FeynDiag}
\end{figure}

In this paper, we investigate the collider prospects of pair-produced $B$ quarks that decay predominantly to the singlet  $\Phi$ in fully hadronic final states (see Fig.~\ref{fig:FeynDiag}). Earlier, we selected those channels because a top quark can decay to produce a (high-$p_T$, in these cases) lepton in the final states, simplifying the analyses. Because of the large QCD background in a hadron collider, fully hadronic final states are highly complicated to isolate. With sophisticated tagging algorithms, one can still tag a highly boosted hadronic top quark. However, without identifiable known structures, probing a $B$ quark becomes formidable with a simple analysis when the $\Phi$ decays to $bb$ or $gg$ final states. We use a graph neural network (GNN) model to separate the signal events from those coming from the significant background processes after passing them through some basic signal selection criteria. A key advantage of representing event data as graphs is that GNNs are permutation invariant, allowing us to train the model with variable numbers of nodes and edges. Therefore, we can use all the well-defined reconstructed objects in events passing, even those playing no role in the basic selection.

The plan of the paper is as follows. In Section~\ref{sec:VLQ_models}, we sketch an outline of the extended VLQ models. Section~\ref{sec:colliderpheno} describes the event-selection criteria and background processes considered in our analysis. Section~\ref{sec:gnn} presents the GNN model in detail, along with the design choices we made and the training details. In Section~\ref{sec:results}, we present the HL-LHC reach of the singlet and doublet VLQ models. We conclude the paper in Section~\ref{sec:conclu}.   

\begin{figure*}[htp!]
\captionsetup[subfigure]{labelformat=empty}
\subfloat[\quad\quad(a)]{\includegraphics[width=0.75\columnwidth]{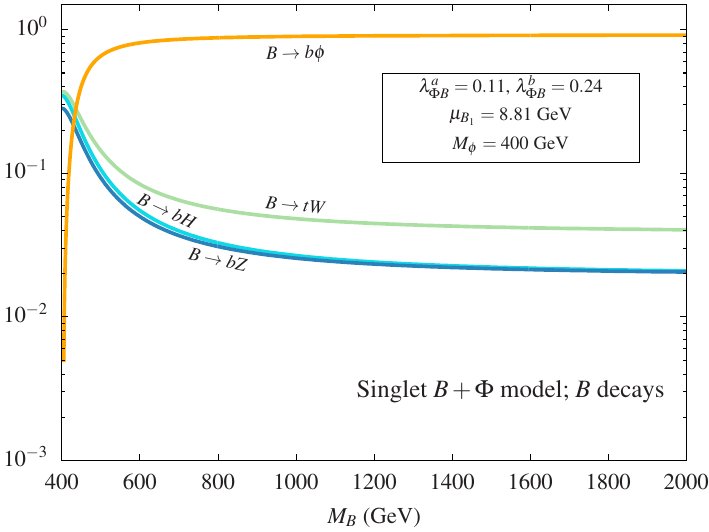}\label{fig:Sing_bMark1_B}}\hspace{0.5cm}
\subfloat[\quad\quad(b)]{\includegraphics[width=0.75\columnwidth]{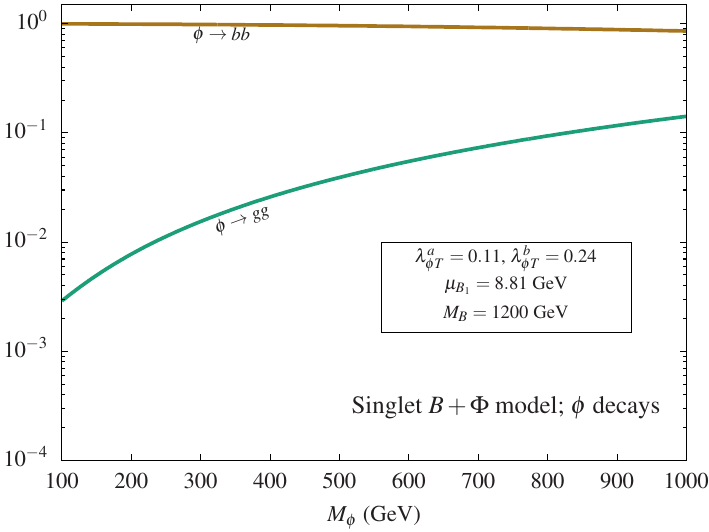}\label{fig:Sing_bMark1_phi}}\\
\subfloat[\quad\quad(c)]{\includegraphics[width=0.75\columnwidth]{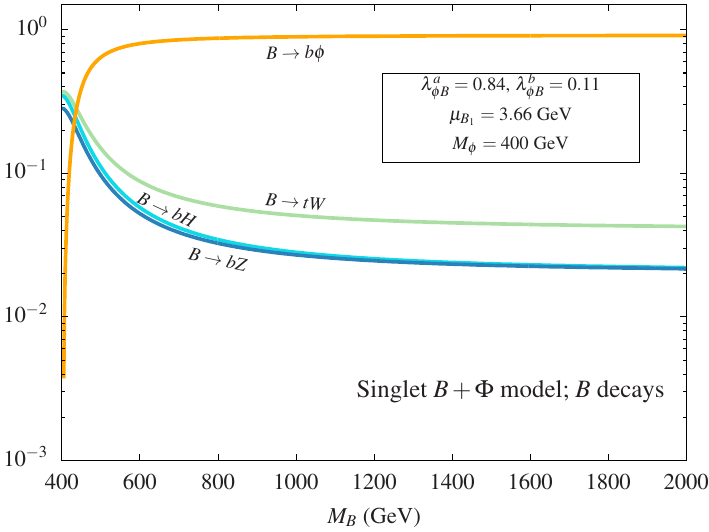}\label{fig:Sing_bMark2_B}}\hspace{0.5cm}
\subfloat[\quad\quad(d)]{\includegraphics[width=0.75\columnwidth]{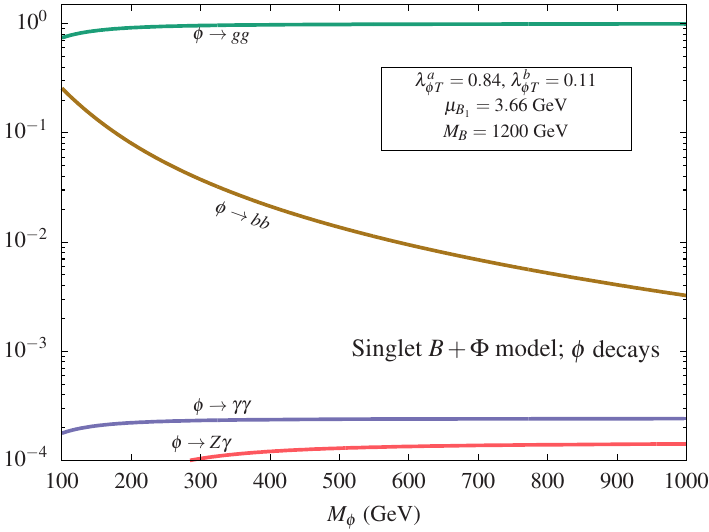}\label{fig:Sing_bMark2_phi}} \\
\subfloat[\quad\quad(e)]{\includegraphics[width=0.75\columnwidth]{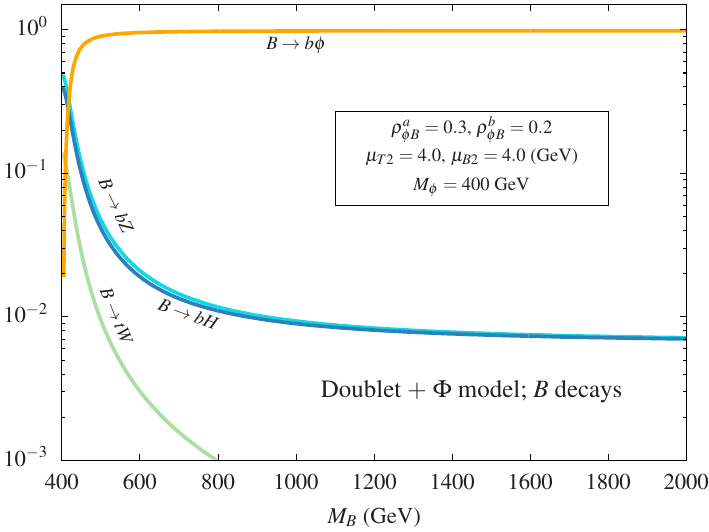}\label{fig:Doub_bMark1_B}}\hspace{0.5cm}
\subfloat[\quad\quad(f)]{\includegraphics[width=0.75\columnwidth]{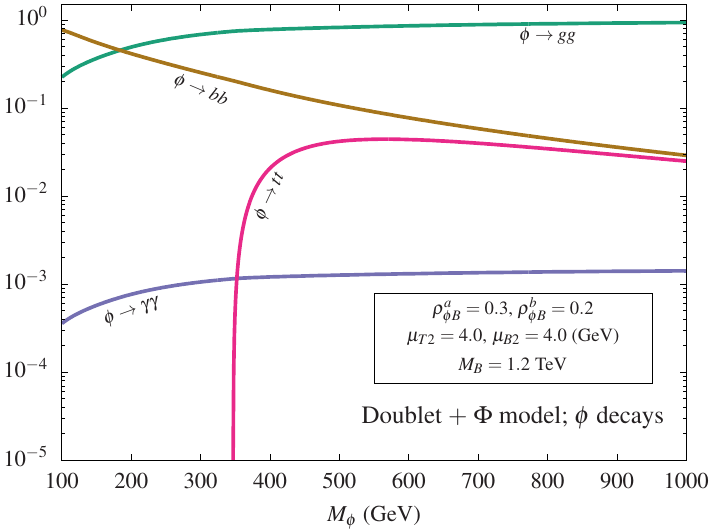}\label{fig:Doub_bMark1_phi}} \\
\subfloat[\quad\quad(g)]{\includegraphics[width=0.75\columnwidth]{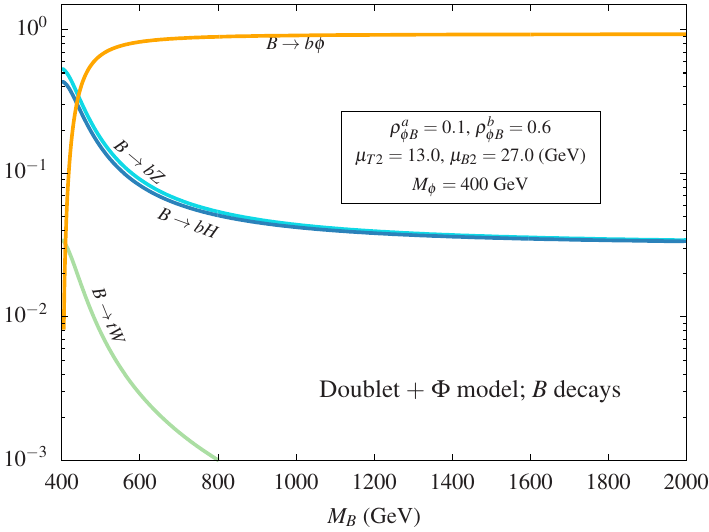}\label{fig:Doub_bMark2_B}}\hspace{0.5cm}
\subfloat[\quad\quad(h)]{\includegraphics[width=0.75\columnwidth]{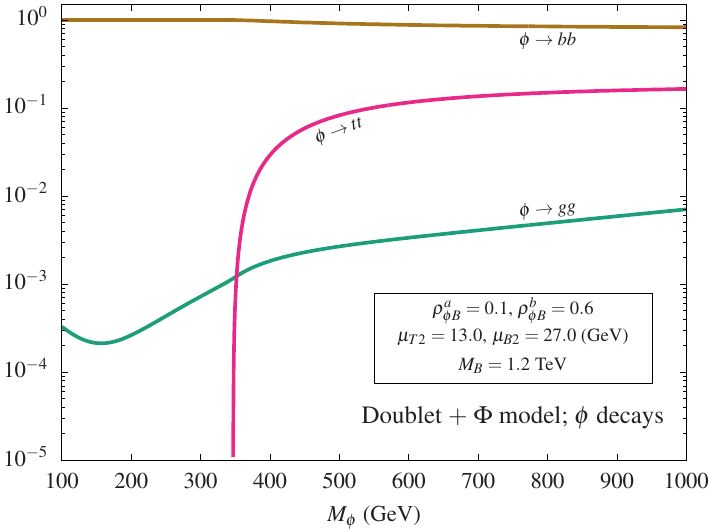}\label{fig:Doub_bMark2_phi}}
\caption{Decays of $B$ and $\phi$ in the singlet $B$ [(a)--(d)] and doublet models [(e)--(h)] for some benchmark parameters where $B\to b\phi$ decay dominates over all the other decay modes. On the right panel, we show that either $\phi\to gg$ or $\phi\to b\bar{b}$ can dominate in both scenarios. For more benchmarks and a detailed discussion on the final states from extended VLQ models, see Ref.~\cite{Bhardwaj:2022nko}.}
\label{fig:br}
\end{figure*}
\begin{figure*}[t!]
\captionsetup[subfigure]{labelformat=empty}
\subfloat[\quad\quad(a)]{\includegraphics[width=0.64\columnwidth]{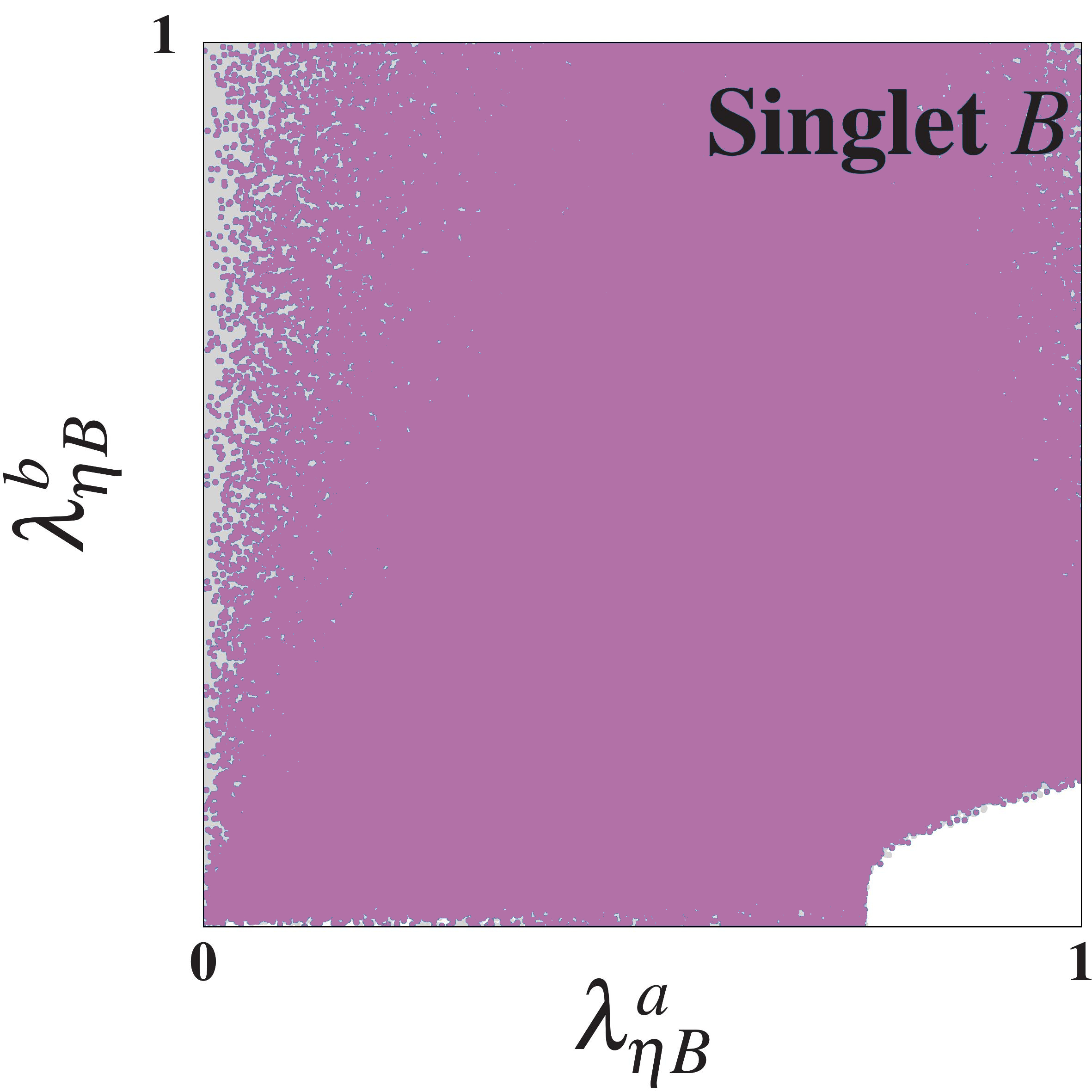}\label{fig:eta_singletB_scan}}\hfill
\subfloat[\quad\quad(b)]{\includegraphics[width=0.64\columnwidth]{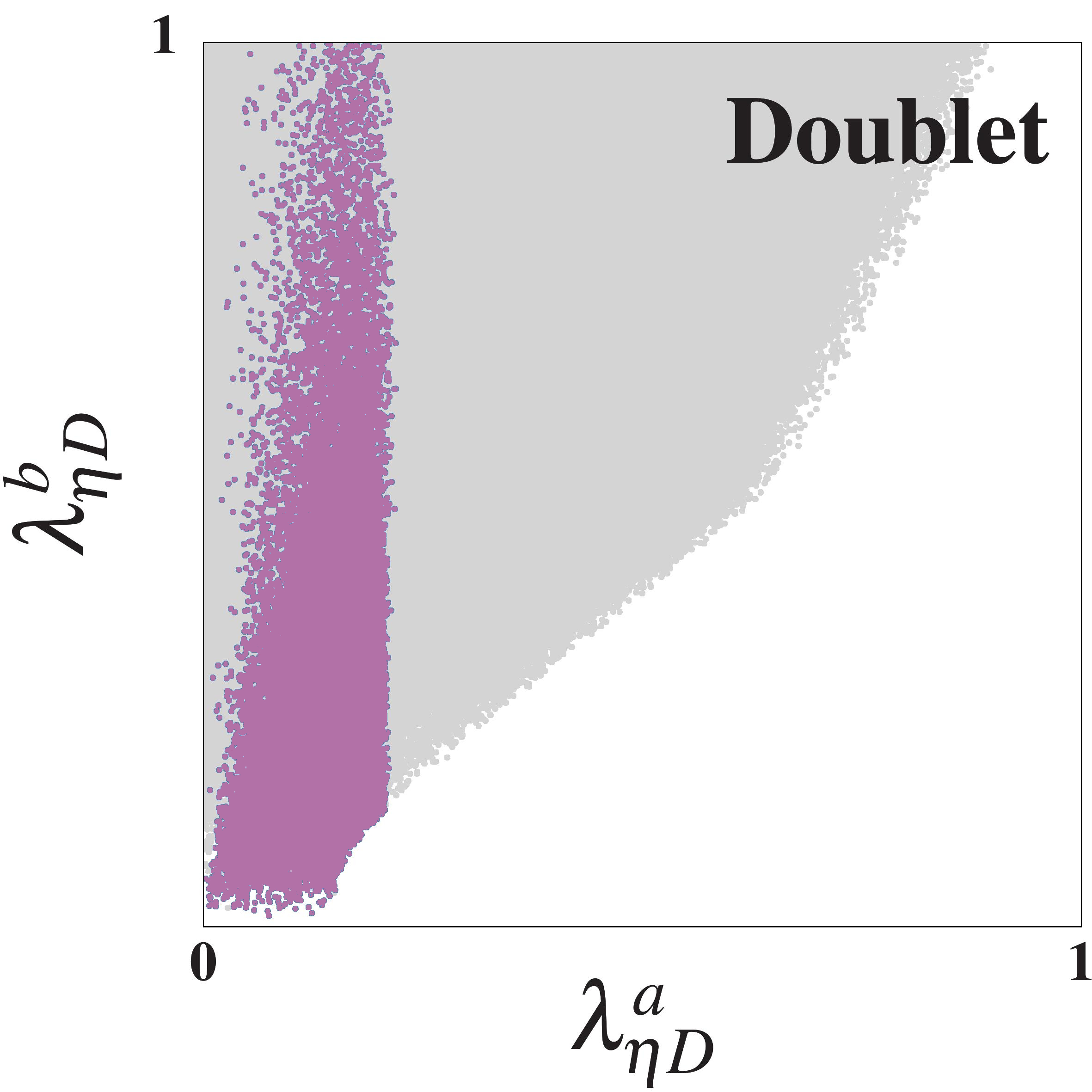}\label{fig:eta_doublet_scan}}\hfill
\subfloat[\quad\quad(c)]{\includegraphics[width=0.64\columnwidth]{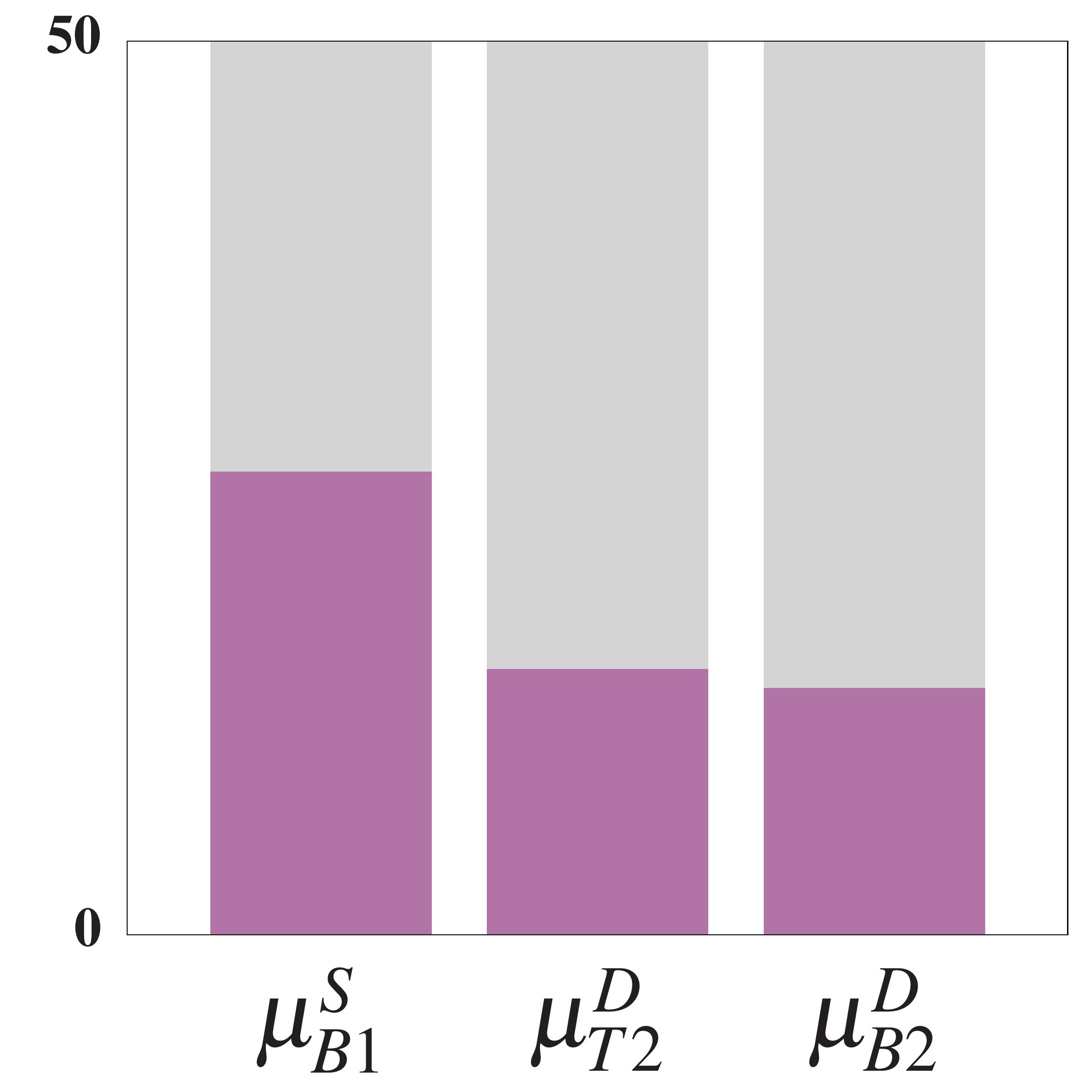}\label{fig:eta_model_mixing}}\\
\subfloat[\quad\quad(d)]{\includegraphics[width=0.64\columnwidth]{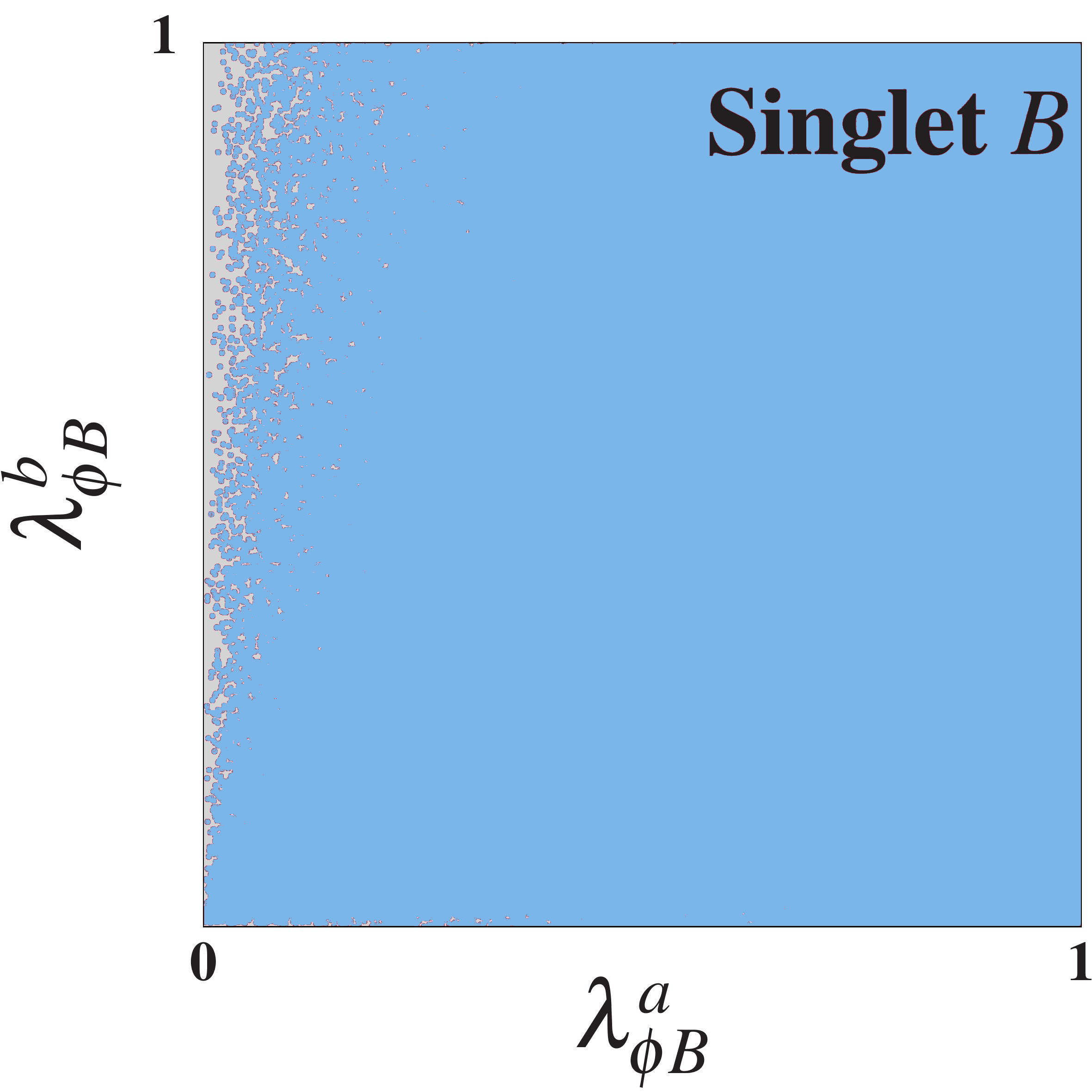}\label{fig:phi_singlet_scan}}\hfill
\subfloat[\quad\quad(e)]{\includegraphics[width=0.64\columnwidth]{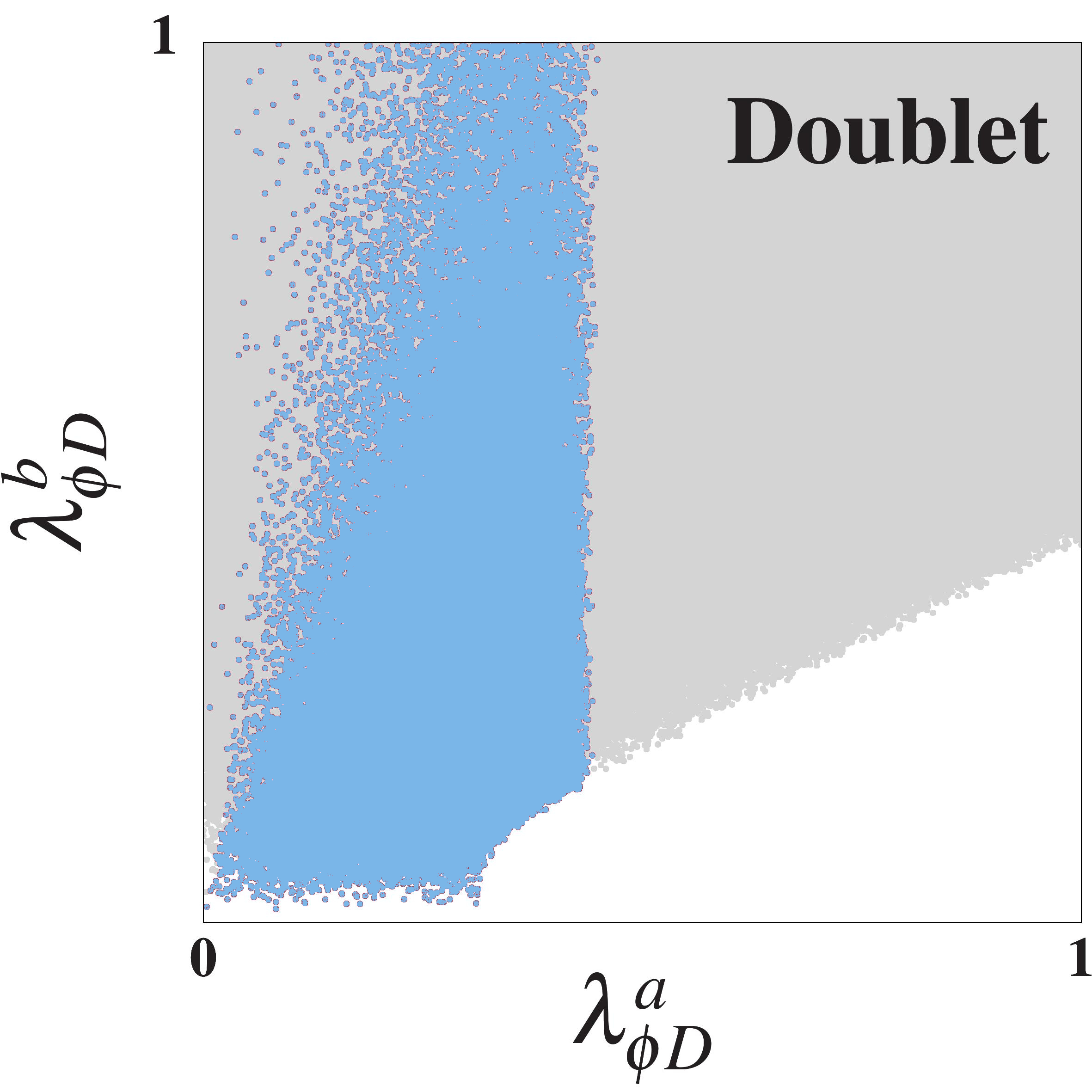}\label{fig:phi_doublet_scan}}\hfill
\subfloat[\quad\quad(f)]{\includegraphics[width=0.64\columnwidth]{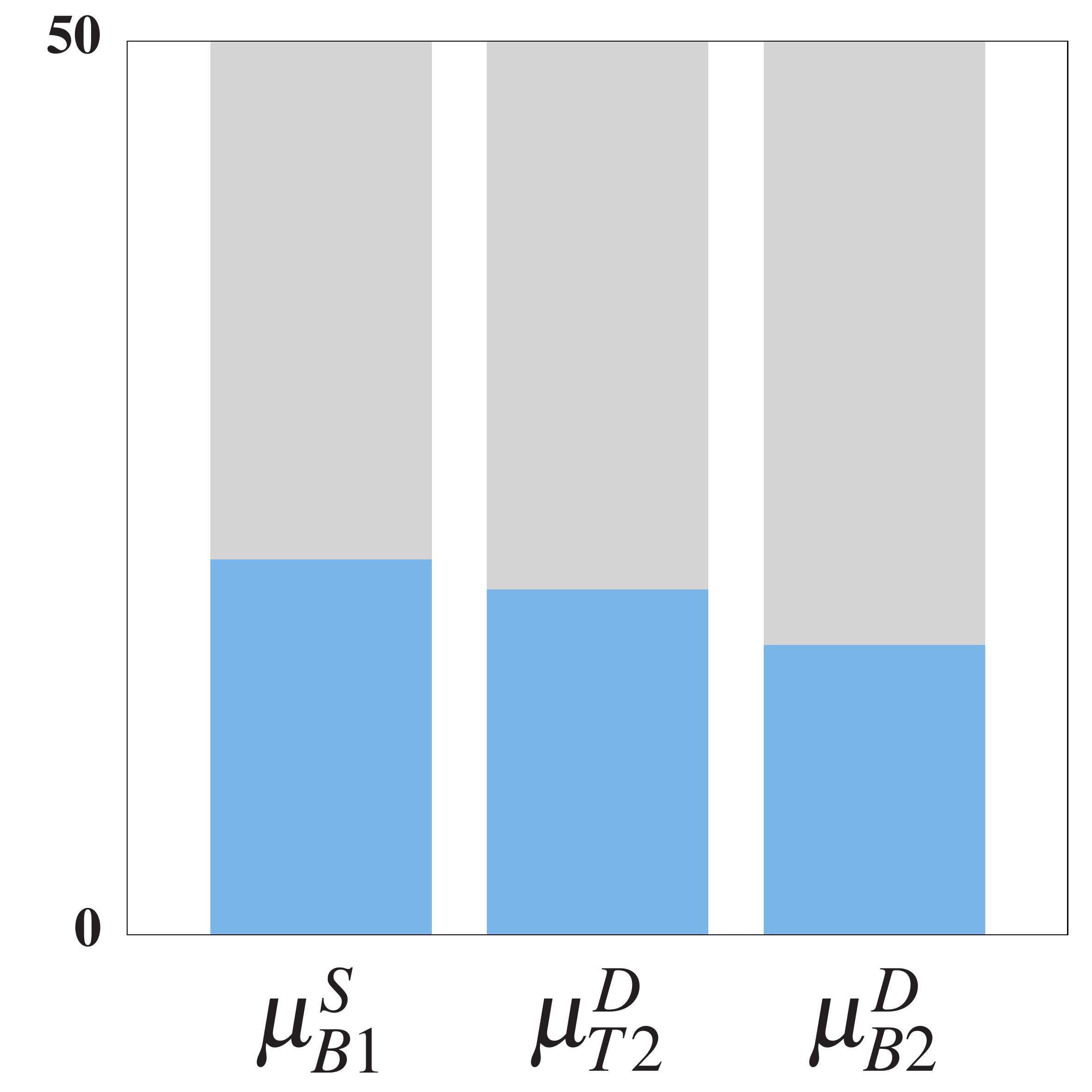}\label{fig:phi_model_mixings}}
\caption{Parameter scans for VLQ $B + \:\Phi$ models for benchmark point $\left( M_B, M_\Phi \right) = \left( 1.2, 0.4 \right)$~TeV. The top row [bottom row] shows the scans for a model where the new singlet is a pseudoscalar $\eta$ [scalar $\phi$] boson. The grey-coloured regions are allowed by reinterpreted LHC limits. Coloured in purple [blue] are the regions where BR$(\eta \to gg) \geq 50\%$ [BR$(\phi \to gg) \geq 50\%$].}
\label{fig:ParamScans}
\end{figure*}

\section{Phenomenological Models}\label{sec:VLQ_models}
\noindent
We are interested in the following signatures:
\begin{align}
pp \to B \bar{B} \to \left(b \Phi\right)\:\left(\bar{b} \Phi\right) \to \left\{ \begin{array}{l} \left(b g g\right) \left(\bar{b} g g\right) \\ 
\left(b b \bar{b}\right) \left(\bar{b} b \bar{b}\right) \end{array}\right..
\end{align}
Below, we discuss the essential phenomenological parameters in the singlet and doublet $B+\Phi$ models that can produce these processes.
\subsection{Singlet $B$ Model}\noindent
In the notation of Ref.~\cite{Bhardwaj:2022nko}, the mass terms relevant to the bottom sector can be written as
\begin{align}
    \mathcal{L} \supset -\Big\{&\lm_{b}\left(\bar{Q}_LH\right)b_R
    +\omega_{B}\left(\bar{Q}_LH\right)B_R\ \nonumber\\
    & + M_{B}\bar{B}_LB_R+ h.c.\Big\},\label{eq:massmat}
\end{align}
where $H=1/\sqrt{2} {\begin{pmatrix} 0 & v \end{pmatrix}}^T$, with $v$ being the vacuum expectation value of the Higgs field, $\lm_b$ is the $b$-quark Yukawa coupling, the off-diagonal $\omega_B$ parametrises the mixing between the $b$ and  $B$ quarks, and $M_{B}$ is the mass of $B$. Once EWS breaks, we get the mass terms as
\begin{equation}
    \mathcal{L}_{mass} = \begin{pmatrix} \bar{b}_L & \bar{B}_L \end{pmatrix} 
    \begin{pmatrix}
    \lambda_b \frac{v}{\sqrt2} & \mu_{B1} \\ 0 & M_B
    \end{pmatrix}
    \begin{pmatrix} b_R \\ B_R \end{pmatrix} + h.c.,
\end{equation}
where $\mu_{B1}=\omega_B v/\sqrt2$. The matrix is diagonalised by a bi-orthogonal rotation with two mixing angles $\theta_L$ and $\theta_R$:
\begin{equation}
    \begin{pmatrix} b_{L/R} \\ B_{L/R}\end{pmatrix} = 
    \begin{pmatrix} c_{L/R} & s_{L/R} \\ -s_{L/R} & c_{L/R} \end{pmatrix}
    \begin{pmatrix} b_{1_{L/R}} \\ b_{2_{L/R}} \end{pmatrix},
\end{equation}
where $s_{P}=\sin\theta_{P}$ and $c_{P}=\cos\theta_{P}$ for the two chirality projections, and $b_1$ and $b_2$ are the mass eigenstates. We identify $b_1$ as the physical bottom quark and $b_2$ is mostly the $B$ quark. Hence, we use the notations $B$ and $b_2$ interchangeably.

If we express the $\Phi$ interactions before the breaking of EWS as
\begin{equation}
    \mathcal{L}_{int}^{B \Phi} = -\lambda^{a}_{\Phi}\Phi \bar{B}_L \Gamma B_R - \lambda_{\Phi}^{b}\Phi \bar{B}_L\Gamma b_R + h.c.,\label{eq:lambdas}
\end{equation}
where $\Gamma=\left\{1, i\gamma_{5}\right\}$ for $\Phi=\left\{ \mbox{scalar~}\phi, \mbox{pseudoscalar~}\eta \right\}$, expanding in terms of $b_1, b_2$ we get
\begin{align}
    \mathcal{L}_{int}^{b_2 \Phi} = &-\lambda_{\Phi}^{a}\Phi\left(c_L \bar{b}_{2L} - s_L \bar{b}_L\right)\Gamma\left(c_R b_{2R} - s_R b_R\right) \nonumber \\
                        &-\lambda_{\Phi}^{b}\Phi\left(c_L \bar{b}_{2L} - s_L \bar{b}_L\right)\Gamma\left(c_R b_{R} + s_R b_{2R}\right) \nonumber\\&+ h.c,
\end{align}
after the symmetry breaking.  
Thus, apart from the masses, the couplings $\lambda_a$ and $\lambda_b$ and the mixing angles form the set of parameters of the Singlet $B$ model. As explained in Ref.~\cite{Bhardwaj:2022nko}, we can draw two types of limits on the VLQ+$\Phi$ models from the current LHC data: 1) We can rescale the current mass exclusion limits on VLQs from their pair-production searches to account for the $\Phi$ decay mode, and 2) recast the heavy resonance searches to put constraints on $\Phi$ production at hadron colliders. The strongest limit on $\Phi$ comes from the clean two-photon search data. (See Refs.~\cite{Bhardwaj:2022nko,Bardhan:2022sif} for a detailed discussion on the constraints on the VLQ+$\Phi$ models from the LHC data.) Fig.~\ref{fig:ParamScans} shows different projections of the allowed parameter region after these constraints are considered. The coloured areas are allowed. The dark region is where the loop-mediated $\Phi\to gg$ decay dominates, i.e., BR$(\Phi \to gg)\geq 50\%$, giving a dijet signature at the colliders. In this region, a dedicated collider analysis in fully hadronic modes is necessary to look for VLQ+$\Phi$ models.  

\subsection{Doublet VLQ}
\noi
When the $T$ and $B$ quarks together forms a weak doublet, $\mathcal{F}=\left(T\ B\right)^T=(\mathbf{3},\mathbf{2},1/6)$,
we can write the terms relevant for the quark masses as
\begin{align}
\mathcal{L} \supset&- \lambda_{t}\left(\bar{Q}_L\widetilde{H}\right)t_R + \rho_{T}\left(\bar{\mathcal{F}}_L \widetilde{H}\right)t_R  + \lambda_{b}\left(\bar{Q}_LH\right)b_R\nonumber\\
&+\rho_{B}\left(\bar{\mathcal{F}}_L H\right)b_R + M_{F}\bar{\mathcal{F}}_L\mathcal{F}_R+ h.c.
\end{align}
From this, we get the following mass matrices:
\begin{align}
\mathcal{L}_{mass}^\mathcal{F} =& \bpm \bar{t}_L & \bar{T}_L \epm 
\bpm 
\begin{array}{cc}
\lm_t \frac v{\sqrt2}  & 0 \\ \rh_{T}\,\frac v{\sqrt2} & M_T
\end{array} 
\epm 
\bpm t_R \\ T_R \epm\nn\\
& +\bpm \bar{b}_L & \bar{B}_L \epm 
\bpm 
\begin{array}{cc}
\lm_b\frac v{\sqrt2}  & 0 \\ \rh_{B}\,\frac v{\sqrt2} & M_B
\end{array} 
\epm 
\bpm b_R \\ B_R \epm + {\tr h.c.}\label{eq:massmatd}
\end{align}
The interactions between $\Phi$ and the doublet $\mathcal{F}$ can be written as
\begin{align}
\mc{L}_{int}^{\Phi \mathcal{F}} =& - \lm_{\Phi D}^a\Phi\,\bar{\mathcal{F}}_L\Gamma \mathcal{F}_R - \lm_{\Phi D}^b \Phi\,\bar{\mathcal{F}}_R\Gm Q_L   + h.c.
\end{align}
Expanding, we get the interactions between $\Phi$ and $B$:
\begin{align}
\mathcal{L} \supset & - \lm_{\Ph D}^a \Phi \lt( c_L^{B}\:\bar{b}_{2L}-s_L^{B}\:\bar{b}_{L}\rt)\Gamma\lt( c_R^{B}\:\bar{b}_{2R}-s_R^{B}\:\bar{b}_{R}\rt) \nn\\
                      & +\lm_{\Ph D}^b \Ph \lt(c_R^{B}\:\bar{b}_{2R} - s_R^{B}\:\bar{b}_{R}\rt)\Gamma\lt(c_L^{B}\:\bar{b}_{L} + s_L^{B}\:\bar{b}_{2L}\rt)+ h.c.\label{eq:LagintD}
\end{align}

In Fig.~\ref{fig:br}, we show the decays of the $B$ quark and the scalar, $\phi$, for some benchmark choices of parameters where the $B\to b\phi$ decay dominates over all other decay modes and $\phi$ (essentially) decays hadronically. Note that since no fine-tuning is needed to find such parameter points, one can easily find similar benchmarks for the pseudoscalar, $\eta$. Fig.~\ref{fig:ParamScans} illustrates the dependence of the dominant decay mode of $\Phi$ on the value of the off-diagonal mass term: $\mu_{B1}=\om_B v/\sqrt2$, for the singlet model, and $\mu^D_{T2} = \mu_{T2} = \rho_T \nu /\sqrt{2}$ and $\mu^D_{B2} = \mu_{B2} = \rho_B \nu /\sqrt{2}$, for the doublet model. The $\Phi\to gg$ mode is enhanced for small mixing values, and the tree-level $\Phi\to b\bar b$ mode starts dominating with increasing $\m_{B1}$. In the doublet mode, $\Phi$ can also decay to a $t\bar t$ pair (if kinematically allowed). However, the combined $\Phi\to gg/bb$ mode dominates. 

\section{Collider phenomenology}\label{sec:colliderpheno}
\begin{figure}[t!]
    \centering
    \includegraphics[width=0.375\textwidth]{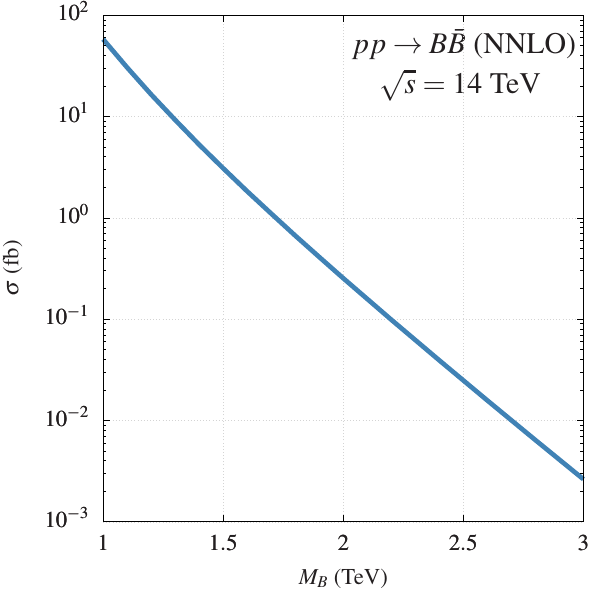}
\caption{The signal cross section at the $14$ TeV LHC calculated from the leading order cross section as $\sg$(LO)$\times K_{\rm NNLO}$. We estimate the NNLO QCD $K$-factor to be about $1.3$ from Ref.~\cite{CMS:2019eqb}. }
    \label{fig:ppXS}
\end{figure}

\noindent 
We use \textsc{FeynRulesv2.3}~\cite{Alloul:2013bka} to obtain the {\sc Universal FeynRules Output}~\cite{Degrande:2011ua} model files, which we use in \textsc{MadGraph5v3.3}~\cite{Alwall:2014hca} to simulate the hard scatterings at the leading order. We use \textsc{Pythia8}~\cite{Sjostrand:2014zea} for showering and hadronisation, and \textsc{Delphes3}~\cite{deFavereau:2013fsa} for simulating a generic LHC detector environment. The events are generated at $\sqrt{s} = 14$ TeV. 
The $b$-tagging efficiency and the mistag rate for the lighter quarks were updated to reflect the Medium working point of the \textsc{DeepCSV} tagger from Ref.~\cite{CMS:2017wtu}. Our analysis relies on two types of jets clustered using the anti-$k_T$ algorithm~\cite{Cacciari:2008gp} --- one with $R=0.4$ (AK4) and the other with $R=1.2$ (fatjet). The AK4 objects are required to pass a minimum-$p_T$ cut: $p_T > 20$ GeV and have $|\eta|<5$. We use the next-to-next-to-leading order (NNLO) signal cross sections from Ref.~\cite{CMS:2019eqb} (Fig.~\ref{fig:ppXS}) and for the background processes, we use the highest-order cross sections available in the literature (Table~\ref{tab:bg-crossx}). We scan over a wide kinematic range --- we take $M_\Phi \in \left[ 300, 1000\right]$ GeV and $M_B \in \left[ 1.0, 3.0 \right]$ TeV.
\begin{table}[b!]
\centering
\caption{Higher-order cross sections of the SM backgrounds considered in our analysis. \label{tab:bg-crossx}}
{\renewcommand\baselinestretch{1.25}\selectfont
\begin{tabular*}{\columnwidth}{l @{\extracolsep{\fill}} lrc}
\hline
Background                   &                   & $\sigma$           & QCD        \\
Processes                    &                   & (pb)               & order      \\
\hline \hline
\multirow{2}{*}{$V$ + jets~\cite{Balossini:2009sa, PhysRevLett.103.082001}} & $W$ + jets  & $1.95 \times 10^5$ & NLO        \\
                                                    & $Z$ + jets  & $6.33 \times 10^4$ & NNLO        \\
\hline
$tt$~\cite{Muselli:2015kba} & $tt$ + jets & $988.57$           & N$^3$LO    \\
\hline
\multirow{3}{*}{Single $t$~\cite{Kidonakis:2015nna}} & $tW$        & $83.10$            & N$^{2}$LO  \\
                             & $tb$              & $248.00$           & N$^{2}$LO \\
                             & $t$ + jets        & $12.35$            & N$^{2}$LO  \\
\hline
\multirow{2}{*}{$VV$ + jets~\cite{Campbell:2011bn}} & $WW$ + jets       & $124.31$           & NLO        \\
                             & $WZ$ + jets       & $51.82$            & NLO        \\
\hline    
\multirow{2}{*}{$VH$~\cite{Cepeda:2019klc}} & $WH$ & $1.50$ & NNLO (NLO EW) \\
                                            & $ZH$ & $0.98$ & NNLO (NLO EW) \\
\hline
\multirow{3}{*}{$ttV$~\cite{Kulesza:2018tqz,LHCHiggsCrossSectionWorkingGroup:2016ypw}}       & $ttZ$             & $1.05$             & NLO + NNLL \\
                             & $ttW$             & $0.65$             & NLO + NNLL \\
                             & $ttH$             & $0.61$           & NLO       \\
\hline                             
\end{tabular*}}
\end{table}


\subsection{The background and selection criteria}
\begin{figure}[!t]
    \centering
    \includegraphics[width=\linewidth]{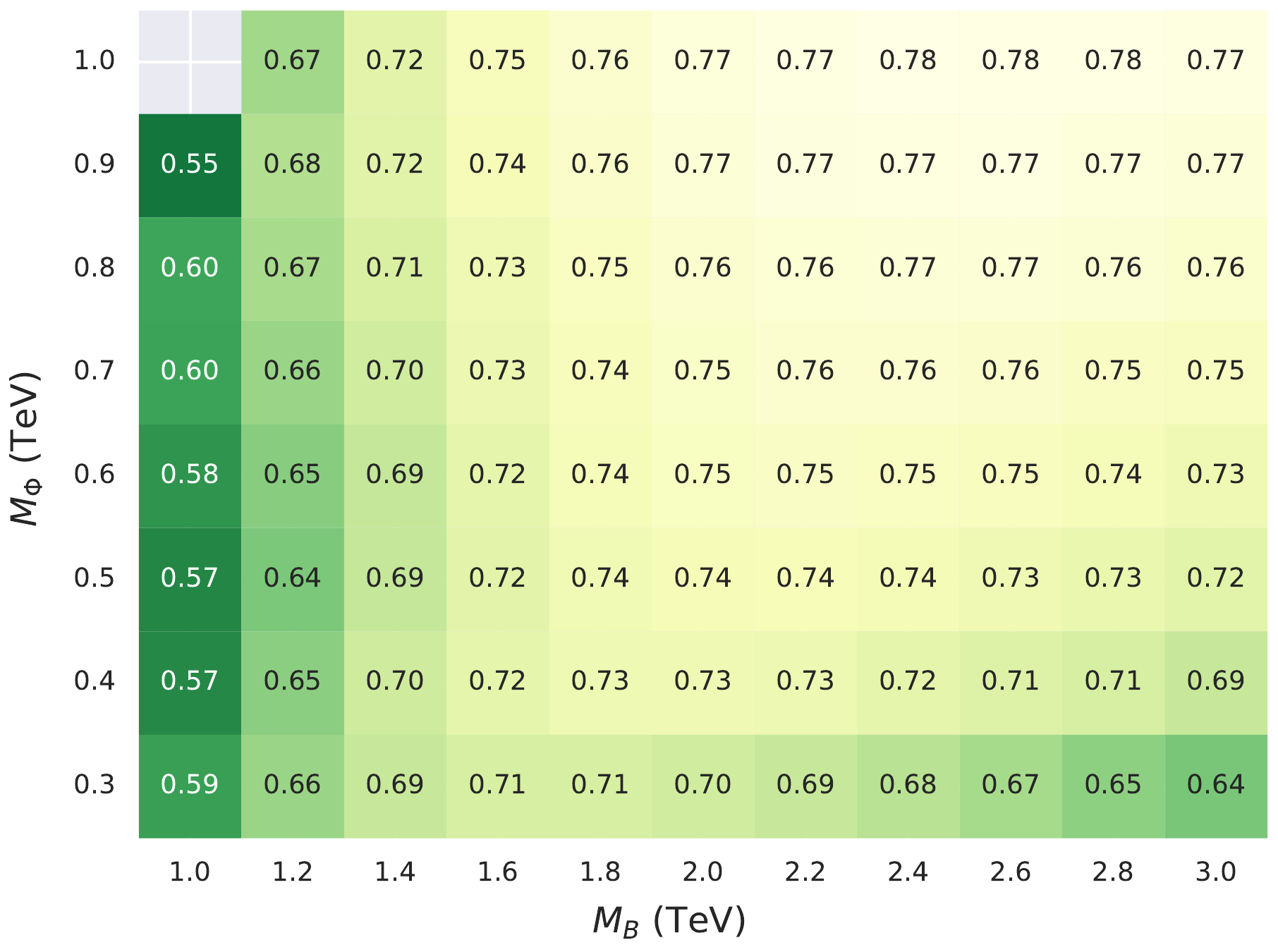}
\caption{Selection efficiency after $\mathfrak C_5$ at various parameter points on the $M_B-M_\Phi$ plane. On average, the efficiency exceeds $70\%$.\label{fig:sig_surviving_ratios}}
\end{figure}

\noindent
Our signal has at least two $b$ jets and two $\Phi$ (fat)jets originating in heavy particles, but no isolated leptons. We collect all the major background processes that could lead to similar final states in Table~\ref{tab:bg-crossx}.
With the signal topology in mind, we first pass the events through the following selection criteria:

\begin{itemize}
\setlength{\itemindent}{-0.5em}
    \item[$\mathfrak{C}_{1}$:]  \emph{$\ell$-veto, where $l \in \{e, \mu\}$ and $H_T$ > 1 TeV.}
    
    We require an event to have no electrons or muons, and the scalar sum of hadronic transverse momenta ($H_T$) should be at least $1$ TeV.

    \item[$\mathfrak{C}_{2}$:] \emph{At least three AK4 jets with $p_{T1},~p_{T2} > 100$ GeV and $p_{T3},\ldots,p_{Tn}>30$ GeV.}
    
    In our analysis, we only consider reconstructed jets with a minimum $p_T$ of $30$ GeV. We also demand the $p_T$ of the top two leading-$p_{T}$ jets should be at least $100$ GeV.
    
    \item[$\mathfrak{C}_{3}$:] \emph{At least two AK4 jets in the event must be b-tagged.}
  
    \item[$\mathfrak{C}_{4}$:] At least one fatjet $(R=1.2)$ with $p_T > 200$ GeV

    We tag one of the $\Phi$-jets in the event as a fatjet of $R=1.2$ with minimum $p_T > 200$ GeV. We also demand that the mass of fatjet, $M_J > 250$ GeV. To cover both $\Phi\to gg/bb$, our analysis relies only on the two-pronged nature of $\Phi$.
    
    \item[$\mathfrak{C}_{5}$:] \emph{At least two b-tagged jets with $\Delta R_{b_i J}\geq 1.2$}. 
  
    We demand that the event has at least two b-tagged jets that are well-separated from the fatjet, i.e., they do not come from the $\Phi$.
\end{itemize}
The strong $p_T$ cuts and the demand for $b$-tagged jets essentially eliminate the QCD multijet background. These topology-motivated cuts also reduce the total background from other processes significantly (roughly by an order of magnitude) while retaining a large part of the signal. We can see the reduction in Table~\ref{tab:cut-flow}. Fig.~\ref{fig:sig_surviving_ratios} shows the efficiency of the selection cuts at different parameter points. However, as expected, the set of background events remains overwhelmingly larger in this hadronic channel even after the cuts. We collect all the signal and background events that pass the selection cuts and feed them to a GNN model for further analysis.
\begin{table*}
\begin{centering}
\caption{Cut flow for the four benchmark choices of the signal and the relevant background processes. We use the MLM jet-parton shower matching technique~\cite{Mangano:2006rw} to generate the background samples as indicated by the additional jets in brackets. Wherever necessary, we simulate background events with generation-level cuts to reduce computational time. The events are estimated for luminosity $\mathcal{L}=3$ ab$^{-1}$. \label{tab:cut-flow}}
{\renewcommand\baselinestretch{1.25}\selectfont
\begin{tabular*}{\textwidth}{l @{\extracolsep{\fill}} 
rrrrr}
\hline
 & \multicolumn{5}{c}{Selection Criteria}\\\cline{2-6} 
                                 & $\mathfrak{C}_{1}$ & $\mathfrak{C}_{2}$ & $\mathfrak{C}_{3}$ & $\mathfrak{C}_{4}$ & $\mathfrak{C}_5$  \tabularnewline
\hline\hline
\multicolumn{6}{c}{Signal benchmarks} \\
\hline
$M_{B}=1200$ GeV, $M_{\Phi}=400$ GeV       
& $\num{4.90e4}$           & $\num{4.90e4}$           & $\num{4.76e4}$           & $\num{3.76e4}$           & $\num{3.22e4}$            \tabularnewline
$M_{B}=1200$ GeV, $M_{\Phi}=900$ GeV       
& $\num{4.91e4}$           & $\num{4.91e4}$           & $\num{4.78e4}$           & $\num{3.90e4}$           & $\num{3.35e4}$            \tabularnewline
$M_{B}=2000$ GeV, $M_{\Phi}=700$ GeV      
& $754$            & $754$            & $725$            & $664$            & $571$           \tabularnewline
$M_{B}=2800$ GeV, $M_{\Phi}=400$ GeV      
& $19$             & $19$             & $18$             & $18$             & $14$  \tabularnewline
\hline

\multicolumn{6}{c}{Background processes} \\
\hline

$t_{h}t_{h}~(+j)$
& \num{2.21e7}	& \num{2.21e7}	& \num{1.22e7}	& \num{4.504e6}	& \num{1.38e6}  \tabularnewline

$Z_{bb} ~(+j)$
& \num{1.36e7}	& \num{1.18e7}	& \num{2.54e6}	& \num{4.11e5}  & \num{1.45e5}  \tabularnewline

$tb~(+j)$
& \num{4.28e5}	& \num{4.14e5}	& \num{2.20e5}	& \num{6.21e4}	& \num{1.44e4}  \tabularnewline

$t_{h} t_{h}H ~(+j)$
& \num{5.50e4}	& \num{5.50e4}	& \num{4.74e4}	& \num{2.40e4}	& \num{1.13e4}  \tabularnewline

$W_h~(+j)$
& \num{8.34e6}	& \num{7.09e6}	& \num{1.85e5}	& \num{3.98e4}	& \num{8.74e3}  \tabularnewline

$t_{h}t_{h}W_{h}~(+j)$
& \num{7.15e4}	& \num{7.15e4}	& \num{3.97e4}	& \num{1.84e4}	& \num{6.67e3}  \tabularnewline

$W_{h} Z_{bb} ~(+j)$
& \num{1.84e5}	& \num{1.81e5}	& \num{6.24e4}	& \num{2.13e4}	& \num{5.39e3}  \tabularnewline

$t_{h} t_{h}Z_h ~(+j)$
& \num{4.58e4}	& \num{4.57e4}	& \num{2.83e4}	& \num{1.28e4}	& \num{5.17e3}  \tabularnewline

$h_{b} W_h ~(+j)$
& \num{6.77e3}	& \num{98}	& \num{< 10}	& \num{< 10}	& \num{< 10}  \tabularnewline

$h_{b} Z_h ~(+j)$
& \num{4.31e3}	& \num{66}	& \num{< 10}	& \num{< 10}	& \num{< 10}  \tabularnewline

\hline 
\multicolumn{2}{c}{ } & \multicolumn{3}{r}{Total number of background events:} & $\num{1.58e6}$\\
\hline\hline
\end{tabular*}}
\end{centering}
\end{table*}

\section{The Graph Neural Network Model}\label{sec:gnn}
\noindent GNNs are a class of permutation-invariant machine learning methods that learn functions on graphs. The functions may be at the node, edge, or graph levels. The workings of GNNs are better understood by thinking of them as message-passing frameworks: GNN modules (layers) pass messages (information) between nodes to let each node learn the embedding that depends on its attributes and its connections to other nodes in the graph. It can be mathematically expressed as~\cite{Shlomi:2020gdn},
\begin{equation}
\hspace{-5pt}\mathbf{x}_i^{(k)} = \gamma^{(k)} \Big( \mathbf{x}_i^{(k-1)}, \bigoplus_{j \in \mathcal{N}(i)} \phi^{(k)}\big(\mathbf{x}_i^{(k-1)}, \mathbf{x}_j^{(k-1)},\mathbf{e}_{j,i}\big) \Big),
\end{equation}
where $\mathbf{x}_i^{(k)}$ denotes the embedding of the $i$th node at the $k$th layer of the GNN, $\mathcal{N}(i)$ is the neighbourhood of $i$, $e_{j, i}$ is the edge (link) from node $j$ to node $i$, $\bigoplus$ denotes a differentiable, permutation invariant function such as sum, max and min, and $\gamma$ and $\phi$ are differentiable functions like multilayer perceptrons. For graph-level tasks (like our classification task), the latent embeddings of each node are aggregated through a permutation-invariant function to give a single prediction. GNNs can be helpful for tasks such as particle tracking and particle flow~\cite{Kieseler:2020wcq}, jet identification~\cite{Qu:2019gqs,Mikuni:2020wpr}, and, in particular, event identification~\cite{Abdughani:2020xfo,IceCube:2018gms,Abbasi:2022ypr}. GNNs are especially effective since they operate much lower than boosted decision trees and deep neural networks (DNNs), allowing them to model features that might be difficult for us to construct or provide. (See Ref.~\cite{Shlomi:2020gdn} for a comprehensive review of GNNs.)

\subsection{Constructing graphs from a collision event}\label{sec:gcons}

\noindent
Even though our event selection criteria focus on three AK4 jets (of which at least two are $b$ jets) and one fatjet, a typical signal event contains more well-defined objects due to the high jet multiplicity of the $B$-quark decay. In principle, once an event passes our selection criteria, all well-defined objects in that event can be used to classify the event using a GNN model. We can also make use of the event-level features. Therefore, to capture the information content in the events, we model them as heterogeneous graphs. 

We represent the kinematic features of jets and fatjets with special abstract nodes --- we call them `shared attributes' or SA nodes --- essentially containing a set of attributes: $\eta$, $\phi$, $m$, and $p_T$. By constructing an abstract common node type for these kinematic features, we enable the model to process these generic features in a unified way and learn functions to model them better. Each jet or fatjet in a selected event has one corresponding SA node. Since our event topology has two kinds of hadronic objects (jets and fatjets), we choose two kinds of auxiliary nodes to represent them. The auxiliary jet node contains features like $b$ tag, \texttt{MeanSqDeltaR} (the mean of $p_T^2$-weighted RMS distance between the constituent and the jet axis), and \texttt{PTD} (a measure of the $p_T$ dispersion of the jet constituents). The auxiliary fatjet node has $N$-subjettiness ratios and girth (mean of the $p_T$-weighted distances of the constituents from the fatjet axis) along with their four-momenta as its attributes. 

The graph also contains two additional nodes --- \texttt{global} and \texttt{CLS}. The \texttt{global}  node contains features of the whole event, such as $H_T$ and $\slashed{E}_T$. The \texttt{CLS} node extracts the embedding for the entire graph (i.e., event) as a vector. This vector is then fed as the input to a fully connected DNN that performs the signal vs. background binary classification. Table~\ref{tab:node_input_feat} lists the attribute content of each node type we construct. For simplicity, we do not specify edge-level attributes and instead let the GNN module learn these from the node features.
\medskip

\begin{figure*}
    \centering
    \includegraphics[width=0.9\linewidth]{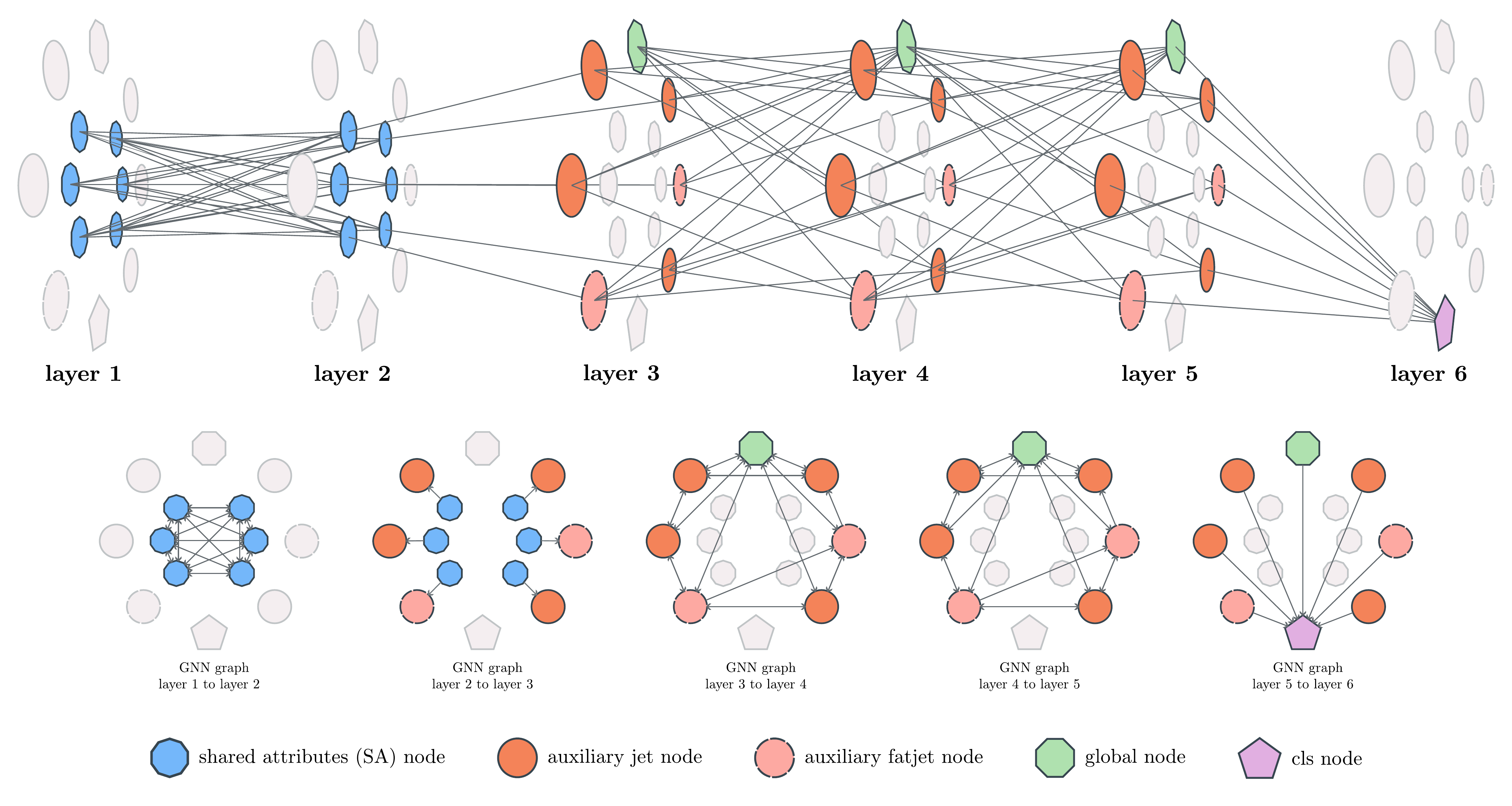}
    \caption{\emph{Constructing the GNN}: We show sequential construction of the graph along with flow of information (message passing) in the GNN model. We use \texttt{GATv2Conv} layer to implement the message passing operation in layer.}
    \label{fig:gnnmoflow}
\end{figure*}

\noindent{\bf Sequential graph contruction:} We sequentially construct the event graph --- Fig.~\ref{fig:gnnmoflow} illustrates the construction.

\begin{enumerate}
    \item[1.] \underline{GNN layer 1-2}: In the first step, the SA nodes are connected with each other (i.e., a clique graph) to enable global interactions among the kinematics of the reconstructed objects and to ensure that each subsequent connection has some information about other nodes present at the global scale. We then perform a round of message passing. 

    \item[2.] \underline{GNN layer 2-3}: Each SA node is connected to its corresponding auxiliary jet or fatjet node in the second step. This allows for object-specific embedding learning, where the kinematic information gathered from the previous layer and the node-level input from the current one are used to construct meaningful embeddings for each reconstructed object. A round of message passing is performed after the connection is made. 

    \item[3.] \underline{GNN layer 3-5}: Jets and fatjets located nearby are likely related. For instance, a fatjet will be close to its constituent jets. Furthermore, independent jets with a common parent are also likely to be close. To incorporate this, we connect nodes to their nearest objects in the $\eta-\phi$ plane in the next step using the $k$ Nearest Neighbours ($k$NN) algorithm. We also connect the \texttt{global} node of the jet and fatjet nodes using a bidirectional link. We do this for two reasons --- we want to provide event-level information to each (fat)jet, and we also want to enable long-range interactions between distant jets, as this information transfer can occur through the \texttt{global} node. This choice is critical for large graphs, and similar strategies have been employed previously~\cite{DBLP:journals/corr/abs-2108-03348,hwang2022vns}. Once the connections are in place, we perform two rounds of message passing. 

    \item[4.] \underline{GNN layer 5-6}: Once all message passings are performed, and each node has an appropriate embedding, we connect all the nodes to the \texttt{CLS} token/node to extract the graph-level class information and perform a round of message passing. This idea is commonly used in Transformers in language-related tasks~\cite{DBLP:journals/corr/abs-1810-04805}.
\end{enumerate}

We primarily use the \texttt{GATv2Conv}~\cite{brody2021attentive} message-passing module to construct our GNN as it shows superior performance over other methods such as \texttt{GCNConv}~\cite{velivckovic2017graph} and \texttt{SAGEConv}~\cite{hamilton2017inductive}. This performance improvement could be attributed to the attention mechanism employed by the GNN. The \texttt{GATv2Conv} module incorporates the attention mechanism to learn an embedding for a node by selectively attending to information obtained by its neighbours,
\begin{equation}
    \mathbf{x}^{\prime}_i = \sum_{j \in \mathcal{N}(i) \cup \{ i \}}
\alpha_{i,j}\mathbf{\Theta}_{t}\mathbf{x}_{j},
\end{equation}
where the attention coefficients $\alpha_{i, j}$ are computed as 
\begin{equation}
    \alpha_{i,j} =
\frac{
\exp\left(\mathbf{a}^{\top}\mathrm{LeakyReLU}\left(
\mathbf{\Theta}_{s} \mathbf{x}_i + \mathbf{\Theta}_{t} \mathbf{x}_j
\right)\right)}
{\sum_{k \in \mathcal{N}(i) \cup \{ i \}}
\exp\left(\mathbf{a}^{\top}\mathrm{LeakyReLU}\left(
\mathbf{\Theta}_{s} \mathbf{x}_i + \mathbf{\Theta}_{t} \mathbf{x}_k
\right)\right)}.
\end{equation}
Here, subscripts $s$ and $t$ indicate the source and the target in the message passing framework, and $\mathbf{\Theta}_{s,t}$ are their corresponding parameters. Using the constructed event embeddings from \texttt{CLS} node, we use a simple, fully connected deep neural network (DNN) to perform the final event classification.

\begin{figure*}
    \includegraphics[width=0.8\linewidth]{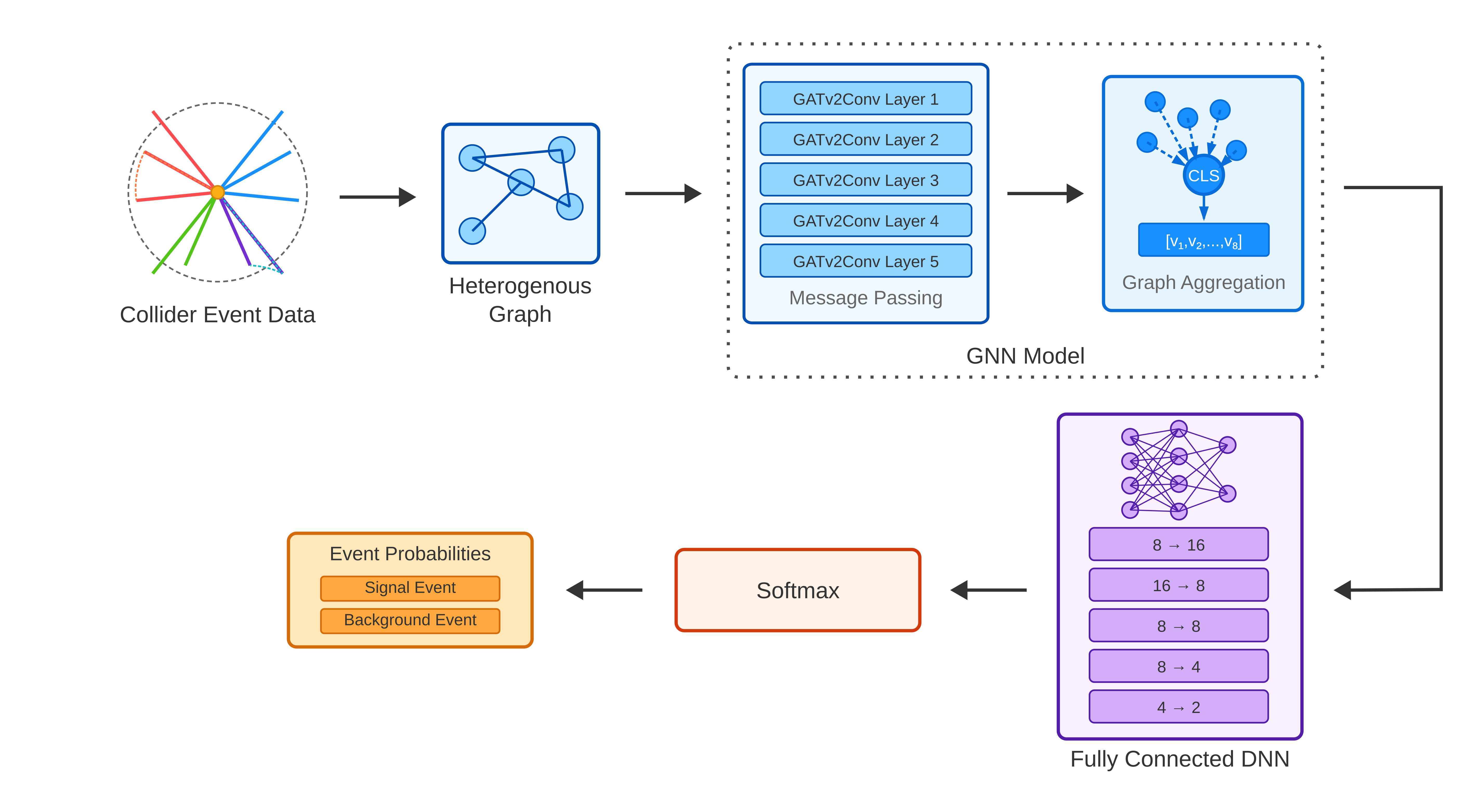}
    \caption{Schematic of the event classification pipeline. We perform an event-level signal vs. background classification on the learned event embeddings using a DNN model. We use a GNN model (Fig.~\ref{fig:gnnmoflow}) to learn the embeddings of the event.}
    \label{fig:pipeline}
\end{figure*}
\begin{table}[]
    \centering
    \begin{tabular*}{\columnwidth}{l @{\extracolsep{\fill}}c}
        \hline
        Node-type & Input Features \\
        \hline \hline
        Shared attributes (SA) nodes & $m, p_T, \eta, \phi$  \\
        Auxiliary jet & \texttt{MeanSqDeltaR}, $b$-tag, \texttt{PTD} \\
        Auxiliary fatjet & girth, $E, p_x, p_y, p_z, \tau_{i}, \tau_{ij}$   \\
        \texttt{global} & $H_T, \slashed{E}_T$ \\
        \texttt{CLS} & -- \\
        \hline \hline
    \end{tabular*}
    \caption{Node types and their respective input features. For the auxiliary fatjet node, $\tau$ denotes the $N$-subjettiness feature, where $i = 1, 2, 3, 4$ and $\tau_{ij}$ denotes the $N$-subjettiness ratios ($j = 1, 2, 3$).}
    \label{tab:node_input_feat}
\end{table}
\subsection{Classifier description and hyperparameters tuning}\label{sec:classifier}

\noindent
Our classifier has two components: a GNN model (with six graph-convolution layers) to learn the event embedding and a DNN model (with five fully connected layers) for event classification using the embeddings. Fig.~\ref{fig:pipeline} shows a representative diagram of the classification pipeline. We use  \texttt{BatchNorm}~\cite{2015arXiv150203167I} and \texttt{Dropout}~\cite{JMLR:v15:srivastava14a} layers to regularise our model with dropout probability $p=0.5$. We use the \textsc{AdamW} optimiser~\cite{2017arXiv171105101L} with a weight decay value of $\lambda = 10^{-4}$.  We run a comprehensive hyperparameter search on the depth of the model (both GNN and DNN), the sizes of the various nodes of the different layers, the regularisation parameters, the learning rate and weight decay, the batch size, the number of epochs of training, and the value of $k$ for $k$NN for generating the graph. Our hyperparameters are optimised using the Bayesian Search strategy on \textsc{Weights and Biases}~\cite{wandb}. 
\begin{figure}
    \centering
    \includegraphics[width=\linewidth]{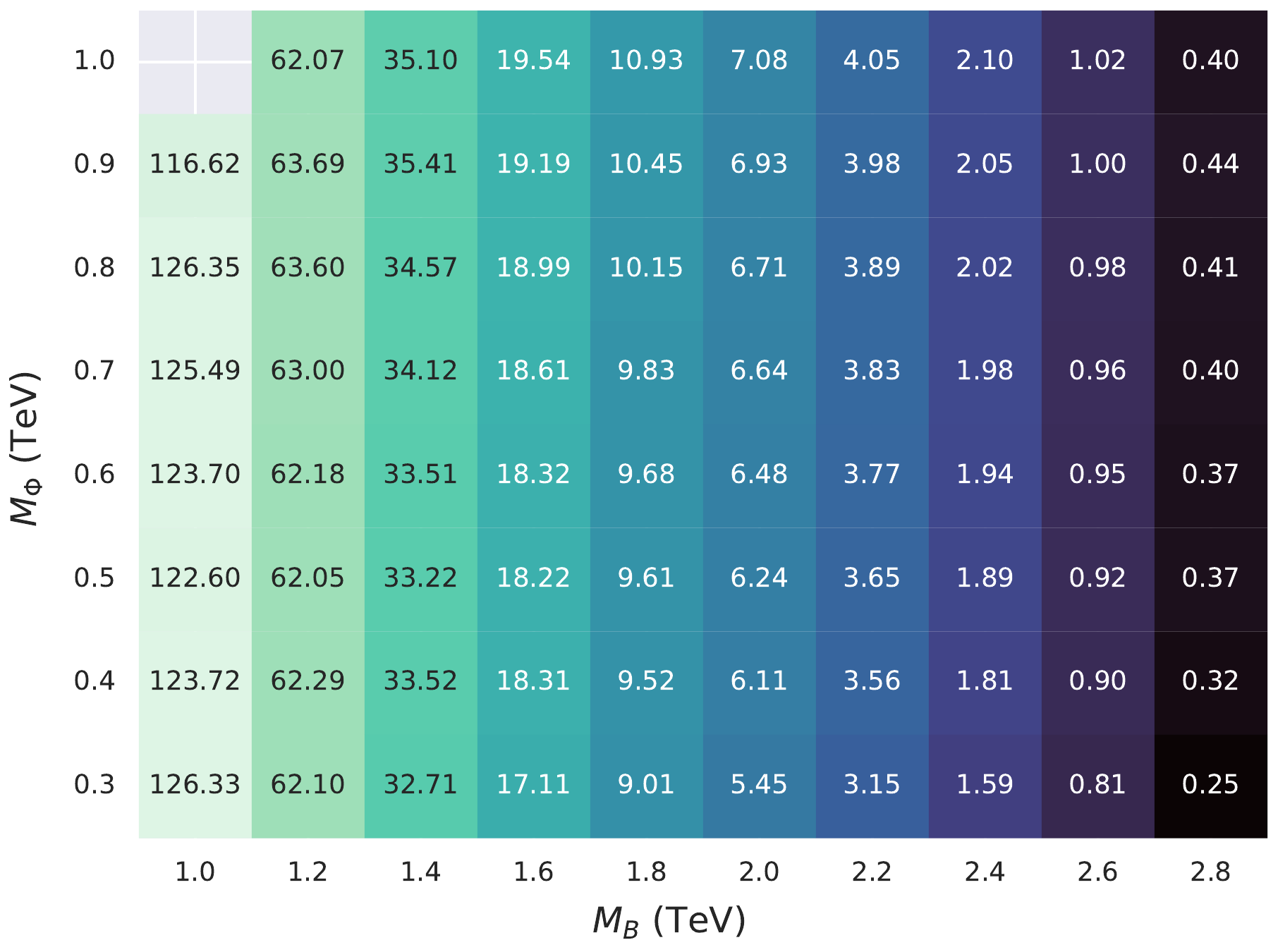}
    \caption{Discovery sensitivity [defined in Eq.~\eqref{eq:z_disc}] obtained from the GNN+DNN model for each parameter point on the $M_B-M_\Phi$  grid considered in Fig.~\ref{fig:sig_surviving_ratios}.}
    \label{fig:GNN_res_hmap}
\end{figure}

\begin{figure*}[]
\captionsetup[subfigure]{labelformat=empty}
\begin{centering}
\subfloat[\quad\quad(a)]{\includegraphics[width=0.42\textwidth]{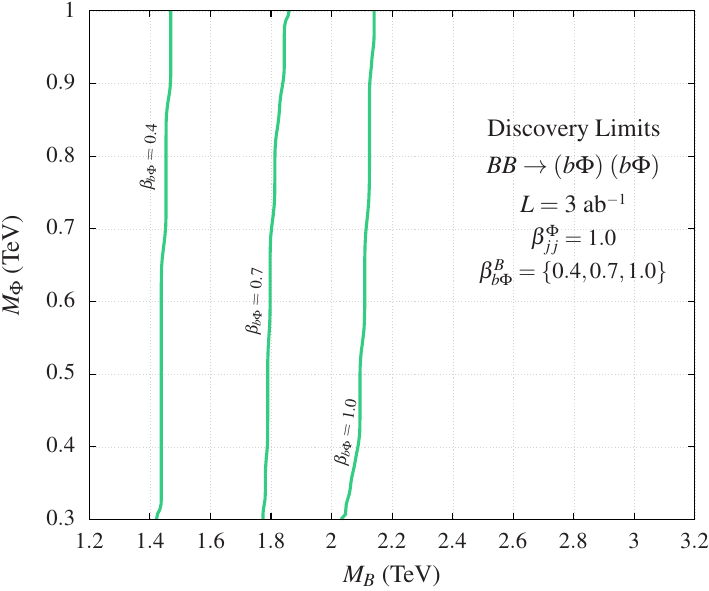}}
\hspace{1cm}\subfloat[\quad\quad(b)]{
\includegraphics[width=0.42\textwidth]{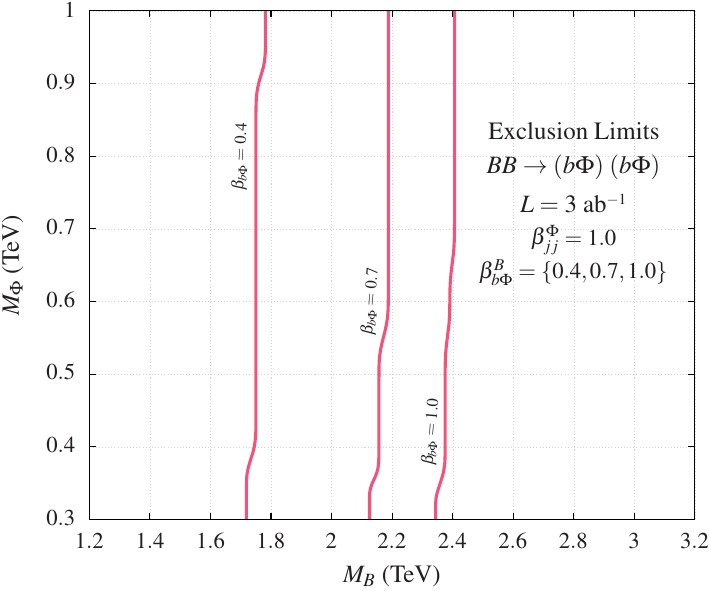}
}\\
\subfloat[\quad\quad(c)]{\includegraphics[width=0.42\textwidth]{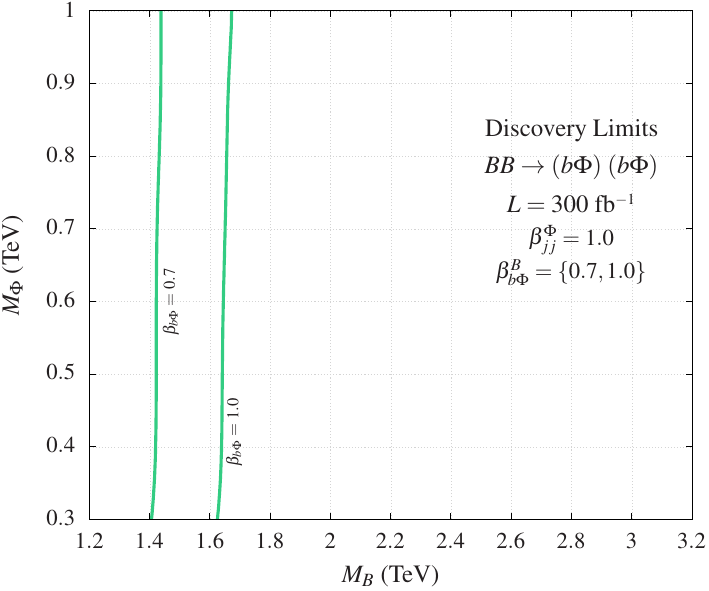}}
\hspace{1cm}\subfloat[\quad\quad(d)]{
\includegraphics[width=0.42\textwidth]{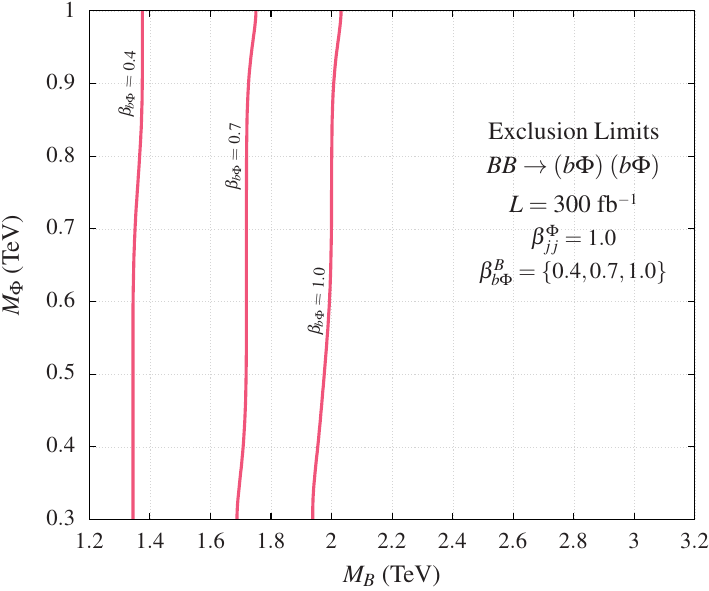}
}

\par\end{centering}
\caption{ 
LHC reach plots for $BB \to \left(b\Phi\right)\:\left(b \Phi\right)$ signal at the HL-LHC, i.e., for luminosity, $L = 3000$ fb$^{-1}$ [(a) and (b)]. The left panel shows the $\mc Z_D=5$ [discovery, Eq.~\eqref{eq:z_disc}] contours and the right panel shows the $\mc Z_E=2$ [$\sim$ exclusion, Eq.~\eqref{eq:z_excl}] contours for three different values of branching in the $B \to b\Phi$ mode: $\beta_{b \Phi} = 40\%, 70\%, 100\%$. The boson, $\Phi$, has a two-prong decay to either a pair of $b$-jets or gluon jets. We also show these plots for $L = 300$ fb$^{-1}$ in (c) and (d).}\label{fig:lhc-reach}
\captionsetup[subfigure]{labelformat=empty}
\begin{centering}
\subfloat[\quad\quad(a)]{\includegraphics[width=0.42\textwidth]{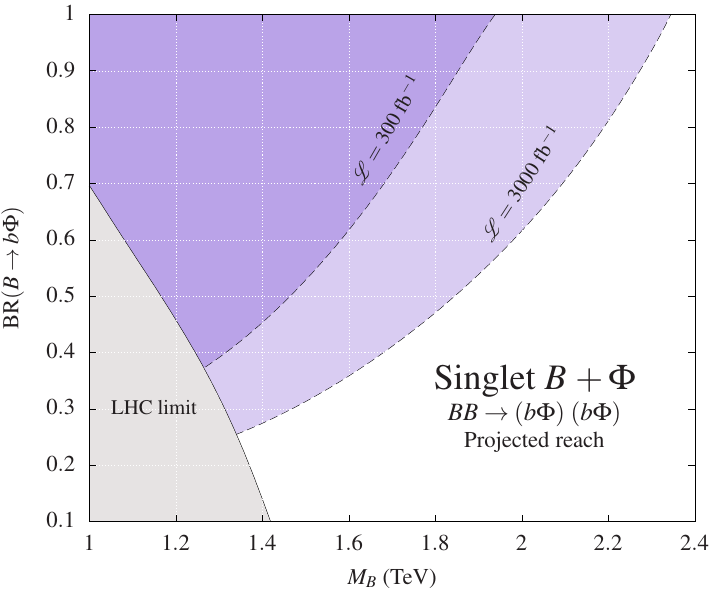}}
\hspace{1cm}\subfloat[\quad\quad(b)]{
\includegraphics[width=0.42\textwidth]{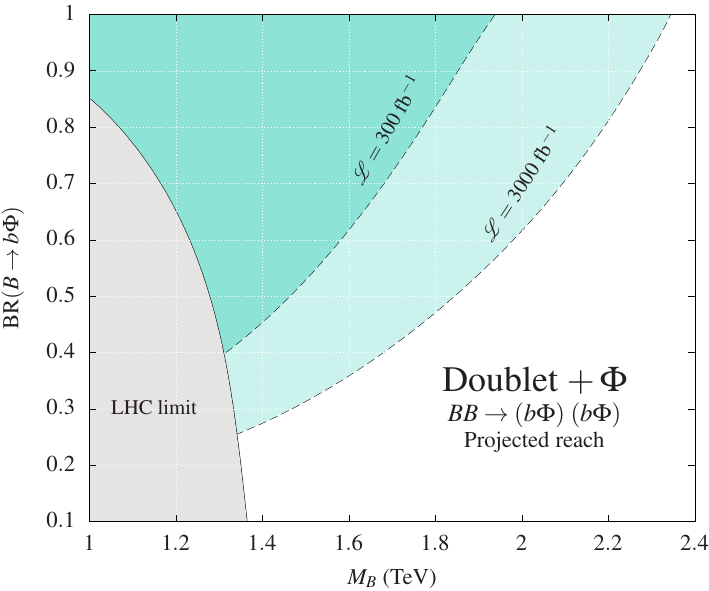}
}\\
\par\end{centering}
\caption{Projected LHC reach for (a) the Singlet $B+\Phi$ model and (b) the Doublet $+\Phi$ in the proposed channel along with the regions ruled out from the current LHC data (in grey) for $L = 300$ and $3000$ fb$^{-1}$.}
\label{fig:model_reach}
\end{figure*}

\subsection{Training Strategy}

\noindent 
We train the classifier with a two-step strategy: 
\begin{enumerate}
    \item \emph{Pre-training}: We pre-train our model on all signal parameter points we consider. 
    \item \emph{Finetuning}: After pre-training the classifier, we finetune our model on each mass point.
\end{enumerate}
Our strategy is mainly motivated by the mounting evidence that large models trained on vast amounts of data generalise better than specific models~\cite{Li:2024htp,Touvron2023LLaMAOA,reid2024gemini,DBLP:journals/corr/abs-2005-14165}. Models trained on larger datasets are better able to capture the underlying properties. 

Our signal is characterised by $M_B$ and $M_\Phi$. We treat the branching ratio of $B$ to the new mode, i.e., BR$(B \to b \Phi)$, as a free parameter that controls the yield of the signal. As mentioned earlier, we pick eight benchmark values for $M_{\Phi}$ (from $300$~GeV to $1$~TeV, sampled at an interval of $100$~GeV), and eleven values for $M_B$ ($1$~TeV to $3$~TeV, sampled at an interval of $200$~GeV). Considering all combinations leaves us with $87$ distinct parameter points (the parameter point $M_B = M_\Phi = 1$ TeV is kinematically forbidden) to pre-train the model on. The model is trained to perform binary classification at these parameter points. 
\medskip

\noindent \textbf{Loss calculation:}\label{sec:los}
We train an unbiased signal vs background classifier in our pre-training stage, i.e., we set the total background weight equal to the total signal weight. For this, we weigh the background processes such that the total background weight is the proportionate sum of the rates (cross-sections). Therefore, the weight of a particular background process is, 
\begin{equation}
    w_{b_i} \propto \frac{\sigma_{b_i}}{N_{b_i}},
\end{equation}
where $\sigma_{b_i}$ is the cross-section of the background process $i$ and $N_{b_i}$ is the number of samples of background process $i$ in the dataset. Similarly, we weigh the signal events. However, since the signal cross-section varies across the parameter range we scan, we ensure that all points contribute equally to the final loss by setting the weights as 
\begin{equation}
    w_{s} \propto \frac{1}{N_{s}},
\end{equation}
where $N_{s}$ is the number of signal samples for a parameter point. As we showed earlier~\cite{Bardhan:2022sif}, a bias-adjusted loss function performs better than the conventional cross-entropy loss. We find that such a loss function effectively balances the performance across all the signal parameter points while cutting out heavy background processes in the current case as well.
\medskip

\noindent \textbf{Finetuning: }\label{sec:finetune} While finetuning to a particular parameter point, we keep the GNN model frozen and retrain only the fully connected DNN. Each parameter point is finetuned independently, giving us $87$ models in total. We scan the threshold values of the classifier response curve and select the best value that maximises the discovery sensitivity~\cite{Cowan:2010js}, given as
\begin{equation}
    \mathcal{Z}_D = \sqrt{2(N_S + N_B) \ln \left(\frac{N_S + N_B}{N_B}\right) - 2N_S},
    \label{eq:z_disc}
\end{equation}
where $N_S$ and $N_B$ are the number of surviving signal and background events at the HL-LHC, respectively. We take the branching ratio in the new mode $\beta_{b\Phi} = 1$. The pre-trained generic model performs well on most parameter points, as seen in Fig.~\ref{fig:GNN_res_hmap}. Fine-tuning shows a modest improvement in the final $\mathcal{Z}_D$ score, with noticeable improvements at the extremums of the parameter points.

\section{Prospects}\label{sec:results}
\noindent Fig.~\ref{fig:lhc-reach} shows the LHC reach obtained using the finetuned GNN model for two different projected luminosities, $\mathcal{L} = 3000$~fb$^{-1}$ [Figs. (a) and (b) in Fig.~\ref{fig:lhc-reach}] and $300$~fb$^{-1}$ [Figs. (c) and (d) in Fig.~\ref{fig:lhc-reach}]. We show the $5\sigma$ (discovery) and $2\sigma$ ($\sim$ exclusion) contours on the $M_B - M_\Phi$ plane with different contours corresponding to different values of $\beta_{b\Phi}$. Since the number of signal events scales with the branching ratio as $\beta_{b\Phi}^2$, these contours can be easily interpreted for other scenarios (e.g., when the standard decay modes dominate or the $B$ quark has additional decay modes) as well by scaling down the $\beta_{b\Phi} = 1.0$ line appropriately. We estimate the discovery contours with the $Z_D$ score [Eq.~\eqref{eq:z_disc}], and the exclusion limits with~\cite{Cowan:2010js},
\begin{equation}
    \mathcal{Z}_E = \sqrt{2N_S - 2N_B\ln \left(\frac{N_S + N_B}{N_B}\right)},
    \label{eq:z_excl}
\end{equation}
where $N_S, N_B$ are the number of surviving signal and background events at the LHC, respectively. In general, we see that with the complex GNN model, it is possible to attain a discovery significance score of $5\sigma$ at values $M_B > 2$ TeV. The reach is only slightly better towards the higher end of $M_{\Phi}$ considered due to the lower boost of $\Phi$-jet; overall, the GNN+DNN model efficiency follows the same trend as Fig.~\ref{fig:sig_surviving_ratios}.

As expected, the LHC can exclude regions beyond its discovery reach. The projected exclusion limits apply to both the singlet $B+\Phi$ and doublet $+\:\Phi$ models. Since the contours in Fig.~\ref{fig:lhc-reach} are insensitive to $M_{\Phi}$, we reinterpret the exclusion contours and plot them on the BR$(B \to b\Phi)$ vs. $M_B$ plane in Fig.~\ref{fig:model_reach} by picking the maximum value $M_B$ allowed at the $2\sigma$ level for each BR$(B \to b\Phi)$ value $\in \left[0.1, 1.0\right]$. The purple (turquoise) regions are excludable in the singlet $B+\Phi$ (doublet $+\:\Phi$) model. The current limits (recast from the LHC data for $L = 139$ fb$^{-1}$) are coloured in grey. The gain is considerable, particularly at the HL-LHC. It also extends beyond the HL-LHC limits for the monoleptonic final states, as obtained in our previous paper~\cite{Bardhan:2022sif}. There, the singlet $B\:+\:\Phi$ exclusion limit stood at $M_B = 1.8$~TeV, for BR$(B\to b\Phi) = 50\%$, corresponding to the maximal signal yield.

\section{Conclusions}\label{sec:conclu}
\noindent In this paper, we studied the pair production of vectorlike bottom-type quarks ($B$), with each decaying to a $b$ quark and a new gauge-singlet scalar or pseudoscalar boson $\Phi$ that dominantly decays to $gg$ or $b\bar{b}$ pairs. Such final states ($2b + 4j$ or $ 6 b$) are difficult to isolate due to the absence of any leptonic handles and an overwhelmingly large hadronic background at the LHC. We designed a sophisticated GNN model-based event classification pipeline to tag the resulting fully hadronic final states. By representing collision events as heterogeneous graphs containing jets, fatjets, and global event features, our model utilised the permutation invariance and message-passing strengths of GNNs. We also employed a novel two-step training strategy to train our event classifier: a pre-training over the entire set of benchmark parameter points before fine-tuning the model for classification at a particular parameter point. This strategy closely follows the one used to train large foundation models in various domains (see, e.g., Refs.~\cite{Li:2024htp,Touvron2023LLaMAOA,reid2024gemini,DBLP:journals/corr/abs-2005-14165}). Our estimation showed that, at the HL-LHC ($L = 3000$ fb$^{-1}$), $M_B$ up to $2.4$~TeV could be excluded by a dedicated search in the joint $2b + 4j/6b$ channel for BR$(B\to b\Phi) \approx 100\%$. Even for a moderate BR in the new mode, $\beta_{b\Phi} = 0.4$, the exclusion limit in $M_B$ goes up to about $1.8$~TeV. Our projections indicate that, in the future, the LHC sensitivity to the singlet and doublet VLQ scenarios (see Fig.~\ref{fig:model_reach}) could be significantly enhanced, particularly in regions of parameter space currently unconstrained. Thus, a dedicated search for pair-produced VLQs decaying through the $4j+2b\:/\:6b$ channel could uncover or constrain a wide range of new physics scenarios. 

Though our analysis is thorough, a more rigorous analysis with real data, higher-order simulations, or better simulation of detector effects (e.g., with Geant4~\cite{GEANT4:2002zbu}) could improve our results. Our estimations can be taken as conservative since, in the signal, we only included events where each pair-produced $B$ quark decayed to a $b$ and a $\Phi$ pair. Including hadronic final states from all pair-production signatures arising from standard decays of the $B$ quark (i.e., $B \to \left\{ tW, bZ, bh \right\}$) that satisfy our selection criteria could enhance the signal strength, as we showed in Ref.~\cite{Bardhan:2022sif} while studying the prospects of an inclusive $tW + X$ signal arising from the decays of pair-produced $B$ quarks. Moreover, as Refs.~\cite{Mandal:2012rx, Mandal:2015vfa} showed, accounting for contaminating contributions from other production processes (e.g., single productions) could further enhance sensitivities. This is interesting because, beyond $2$ TeV, the pair production process may not remain the leading production process of VLQs. Moreover, non-conventional single productions ($pp\to B\Phi$) also come into play in that regime. On the analysis front, transformer-based models have been shown to be effective at HEP tasks~\cite{Qu:2022mxj,Bogatskiy:2023nnw,Finke:2023veq,Golling:2024abg,Wu:2024thh,Bardhan:2025icr}. Though their performance on event-level features or heterogeneous reconstructed objects remains to be studied rigorously, using them in the analysis pipeline could improve the prospects further.

\section*{acknowledgement}
\noindent  T.M. acknowledges partial support from the SERB/ANRF,
Government of India, through the Core Research Grant
(CRG) No. CRG/2023/007031. 

\section*{Model Files}
\noindent 
The {\sc Universal FeynRules Output}~\cite{Degrande:2011ua} files used in our analysis are available at \url{https://github.com/rsrchtsm/vectorlikequarks/} under the name {\tt SingBplusPhi}. 

\def\bibfont{\small}
\bibliography{Bprime_hadronic}{}
\end{document}